\documentclass[12pt]{iopart}

\usepackage{graphicx}
\usepackage{subfigure}
\usepackage[sort&compress,square,numbers]{natbib}
\usepackage{hyperref}

\begin{document}

\title[New Scanning Positron Microbeam]{Defect Imaging and Detection of Precipitates Using a New Scanning Positron Microbeam}

\author{T. Gigl, L. Beddrich, M. Dickmann, B. Rien\"acker, M. Thalmayr, S. Vohburger, and C. Hugenschmidt}

\address{Physik-Department E21 and FRM\,II, Technische Universit\"at M\"unchen, Lichtenbergstra\ss e 1, 85748 M\"unchen, Germany}
\ead{christoph.hugenschmidt@frm2.tum.de}

\begin{abstract}
We report on a newly developed scanning positron microbeam based on \textit{threefold moderation} of positrons provided by the high intensity positron source NEPOMUC.
For brightness enhancement a remoderation unit with a 100\,nm thin Ni(100) foil and 9.6\,\% efficiency  is applied to reduce the area of the beam spot by a factor of 60.
In this way, defect spectroscopy is enabled with a lateral resolution of 33\,$\mu$m over a large scanning range of 19$\times$19\,mm${}^2$. 
Moreover, 2D defect imaging  using  Doppler broadening spectroscopy (DBS) is demonstrated to be performed within exceptional short measurement times of less than two minutes for an area of 1$\times$1\,mm${}^2$ (100$\times$100\,$\mu$m${}^2$) with a resolution of 250\,$\mu$m (50\,$\mu$m). 

We studied the defect structure in laser beam welds of the high-strength age-hardened Al alloy (AlCu6Mn, EN AW-2219 T87) by applying (coincident) DBS with unprecedented spatial resolution.
The visualization of the defect distribution revealed a sharp transition between the raw material and the welded zone as well as a very small heat affected zone. 
Vacancy-like defects and  Cu rich precipitates are detected in the as-received material and, to a lesser extent, in the transition zone of the weld. 
Most notably, in the center of the weld vacancies without forming Cu-vacancy complexes, and the dissolution of the Cu atoms in the crystal lattice, i.e.\ formation of a supersaturated solution, could be clearly identified.

\end{abstract}

\maketitle

%++++++++++++++++++++++++++++++++++++++++++++++++++++++++++++++++++++++++++++++++++
\section{Introduction}
%++++++++++++++++++++++++++++++++++++++++++++++++++++++++++++++++++++++++++++++++++
Crystal defects such as dislocations, precipitates and different species of point defects highly influence or even significantly determine the macroscopic physical properties of all kind of materials.
Therefore, the investigation of the nature and concentration of lattice defects plays a major role for an improved understanding of the material properties, which can be deteriorated, e.g. due to the presence of structural vacancies, or considerably improved by deliberate introduction of defects.

In the zone of welded joints, for example, the local mechanical properties are influenced due to the strong spatial dependent structural changes and production of various defects.   
During the welding process the high heat impact  leads to  local softening, e.g. upon friction stir welding, or in case of most other welding techniques which usually are applied it leads to melting of the material and subsequent cooling. 
In this study, we focus on laser beam welds of the age-hardened Al alloy (AlCu6Mn) where a high spatial resolution is needed to study the defect distribution due to the very small welding area. 

Positron annihilation spectroscopy (PAS) has become a well-established tool to investigate lattice defects in solids due to the outstanding sensitivity of positrons to vacancy-like defects \cite{Wes73,Sch88,Pus94,Dupasquier1995}.
After implantation in a material the positron thermalizes rapidly ($\sim$ps), diffuses through the crystal lattice ($\sim$\,100\,nm), and finally annihilates as a delocalized positron from the Bloch-state or from a localized state present in open volume defects. 
The positron-electron annihilation in matter is completely dominated by the emission of two 511\,keV photons in opposite direction in the center-of-mass system.
In the lab system, the transverse momentum of the electron (the momentum of the thermalized positron is negligible) leads to a deviation of the $180\,^{\circ}$ angular correlation and 
the longitudinal projection of the electron momentum onto the direction of the $\gamma$-ray emission $p_L$ results in a Doppler-shift of the 511\,keV $\gamma$-quanta by $\Delta E\,=\,\pm\frac{1}{2}p_Lc$ (see e.g.\,reviews on positron physics \cite{Wes73,Sch88}).
The lower annihilation probability of positrons trapped in vacancies with high-momentum core electrons leads to a smaller $\Delta E$ compared to the bulk lattice. 
In Doppler broadening spectroscopy (DBS) the annihilation photo peak is usually characterized by the S-parameter defined as the ratio of counts in a fixed area around the maximum and the total counts of the annihilation line.
Hence, in a vacancy-like trapped state the S-parameter is enhanced compared to the defect-free state (see, e.g.\,\cite{Hau79}).

In coincident DBS (CDBS) the intrinsic low background allows the detection of events in the outer tail of the 511\,keV photo line, which correspond to the annihilation of inner shell electrons and hence contain valuable information of the chemical surrounding at the annihilation site \cite{Lyn79,Aso96}.
CDBS was applied in a number of experiments to study the chemical surrounding of crystal defects in semiconductors \cite{Ala95,Szp96,Kur98b},  in metals \cite{Par05,Tak06,Sta06}, and precipitates in metallic alloys \cite{Nag01,Nag02}.

Monoenergetic positron beams allow for depth-dependent defect investigations of samples or thin layers.
The positron diffusion length can be measured by variation of the implantation energy that in turn allows to evaluate high vacancy concentrations up to the percent range \cite{Hug15a}.
DBS with energy variable positron beams is particularly suited to investigate the depth distribution of defects. 
There are numerous studies on irradiation induced defects e.g. in Si \cite{Fuj93, Fuj96, Eic97, Bru00}, in oxides \cite{Ued03, Che04}, in polymers \cite{Kob95} or in metals \cite{Tri82, Lyn86, Par15} 
as well as on thin film insulators and semiconductors \cite{Veh87, Nie87, Ued02, Ued16} and thin metal layers  \cite{Ciz09, Eij11}.
By using CDBS, additional elemental information is gained for the investigation of e.g. the oxygen termination of the Si surface \cite{Hug09a}, annealing and alloying of  Au/Cu binary layers \cite{Rei14a} as well as buried metal layers and clusters \cite{Hug08a,Pik11}.
However, using a \textit{scanning} positron beam allows to gain laterally resolved information in two dimensions, and the third dimension is given by the positron implantation energy. 
Besides imaging of the defect distribution in 2D (see e.g.\,\cite{Dav01, Eic07, Hug09b, Rei15a}), 3D imaging was demonstrated on an irradiated quartz sample by Oshima et al.\,\cite{Osh09} and further applied on ion irradiated alloys \cite{Sta07b, Hen12}.
Due to the high intensity provided by the neutron induced positron source NEPOMUC spatially resolved DBS in combination with CDBS can be applied within short measurement times e.g. for the detection of Fe clusters in Al in friction stir welded materials \cite{Hai10} or for the determination of the oxygen deficiency in superconducting thin film oxides \cite{Rei15a}.

For many positron beam applications in solid state physics and materials science a high spatial resolution is extremely beneficial. 
Despite the efforts made, the availability of positron microbeams, if at all,  is very limited due to its highly demanding realization.
However, various proposals have been made for  the development of positron beams providing beam spot diameters of less than 0.1\,mm.
Three decades ago, it was demonstrated that a focused spot size between 12-50\,$\mu$m is achievable with a 10\,keV positron beam \cite{Bra88b}. 
In the late 1990's a positron beam for DBS was set up yielding a diameter of 20-37\,$\mu$m at Bonn university \cite{Gre97}.
At Munich a scanning positron microscope for lifetime measurements has been developed providing a diameter of 20\,$\mu$m \cite{Tri97b} which could be improved to 5\,$\mu$m with a count rate of 100\,cps and, at the expense of count rate, further reduced to about  2\,$\mu$m \cite{Dav01}. 
In Japan, a positron microbeam with a diameter of 80\,$\mu$m and count rate of 50\,cps \cite{Mae07} was improved to a beam diameter of about 4\,$\mu$m with 30-40\,cps for positron lifetime or DB measurements \cite{Mae08}.

Fujinami et al. applied a 150\,nm thick Ni(100) remoderation foil (4.2\,\% efficiency) to produce a $^{22}$Na based positron microbeam with a diameter of 60\,$\mu$m and $<$10\,cps \cite{Fuj08}.
The same technique has been applied at a LINAC based slow positron beam in order to enhance the count rate to 3000\,cps for lifetime measurements with a beam diameter of 90\,$\mu$m \cite{Osh08}.

Apart from the fact that most of these setups are not in (routine) operation, besides the last-mentioned, all microbeams have in common an intrinsically low count rate leading in turn to high measurement times.
In order to enable (C)DB spectroscopy with a spatial resolution well below 100\,$\mu$m, we have developed a new positron microbeam providing high enough intensity and hence reasonable measurement times.
For this purpose, a three-fold moderated positron beam is realized at the upgraded CDB spectrometer located at the high intensity positron source NEPOMUC.

In this paper, we describe the principle, the layout and the performance of the new positron microbeam. Various additional modifications have been made to the CDB spectrometer in order to improve the stability of the beam position and to monitor the beam with high resolution.
Finally, first spatially resolved CDB measurements on  a laser beam weld of a high-strength age-hardened Al alloy (EN AW-2219 T87, AlCu6Mn)  are presented  in order to highlight the excellent capabilities of the new scanning positron microbeam.

%++++++++++++++++++++++++++++++++++++++++++++++++++++++++++++++++++++++++++++++++++
\section{The positron microbeam at the new CDB spectrometer}
%++++++++++++++++++++++++++++++++++++++++++++++++++++++++++++++++++++++++++++++++++
\subsection{Experimental requirements}
The new positron microbeam has been designed  as integral part of the CDB spectrometer upgrade at the positron beam facility NEPOMUC in order to improve considerably the spatial resolution for both, imaging of defect distributions with DBS and defect spectroscopy with CDBS.
By variation of the beam energy, (C)DB measurements can be performed from the surface to a (material dependent) depth of up to a few $\mu$m.
In addition, positron beam experiments should be feasible at elevated temperature (e.g.\,for the in-situ observation of defect annealing) and low temperature (e.g.\,for the investigation of the formation and emission of cold positronium). 
Thus, in order to increase the overall performance of the instrument the following requirements have to be met: 
 
\begin{itemize}
	\item  Providing a beam spot at the sample position of $<$100\,$\mu$m
	\item  Achieving a final implantation energy of up to 30\,keV 
	\item  2D-scanning of a sample area of $\geq$\,15$\times$15\,mm$^2$ with high spatial resolution and within acceptable measurement time
	\item  Adapting the optics for both, the high-intensity \textit{primary} beam (kinetic energy typically 1\,keV) and the low-energy 20\,eV \textit{remoderated} beam provided by NEPOMUC \cite{NEPupg}
	\item  Optimizing the beam guidance and monitoring of the beam parameters (radial position and spot shape) at different positions inside the instrument
	\item  Enabling a wide temperature range between 50\,K and 1000\,K
\end{itemize}

%++++++++++++++++++++++++++++++++++++++++++++++++++++++++++++++++++++++++++++++++++
\subsection{The new CDB spectrometer}
\begin{figure}[htb]
	\centering
\includegraphics[width=\textwidth]{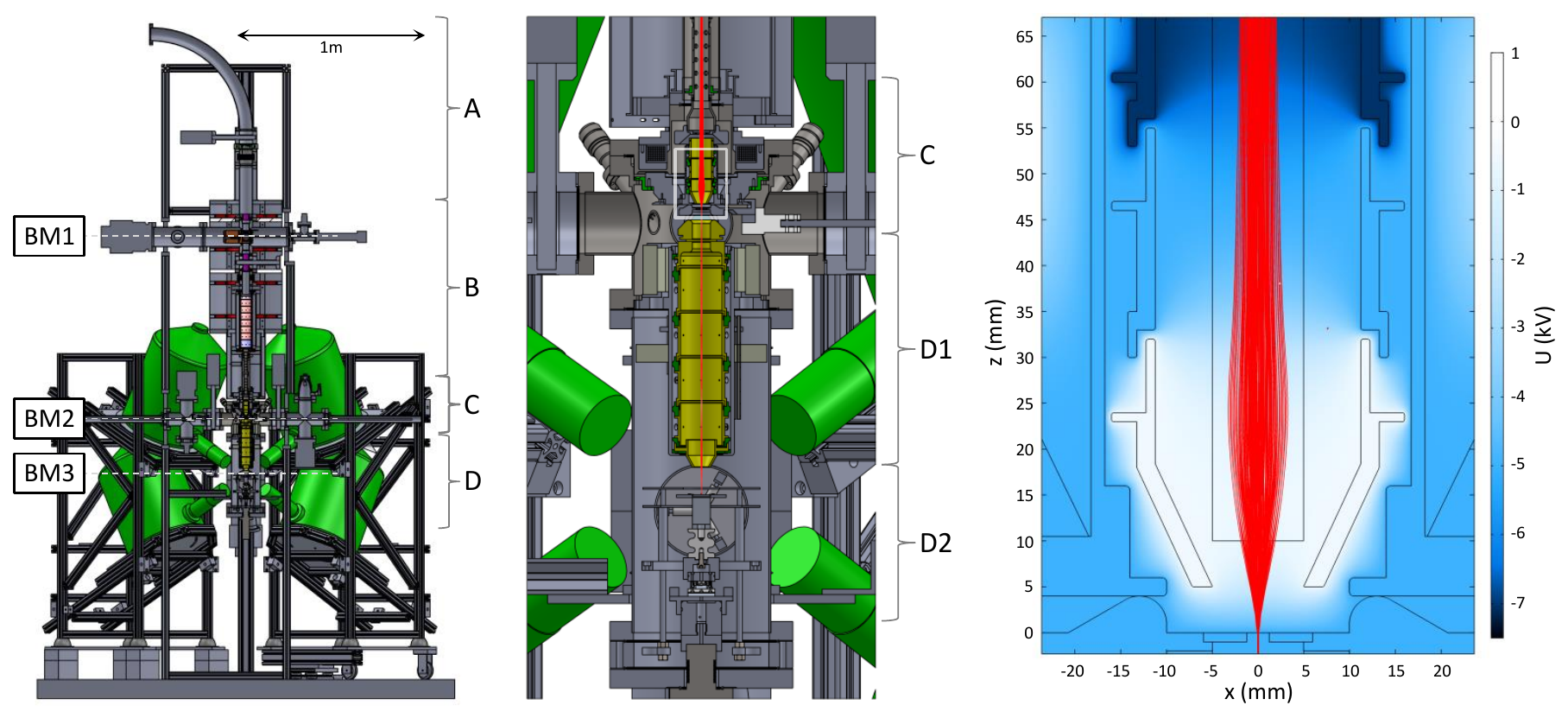}
		\caption{Design of the CDB spectrometer with positron microbeam. 
		Left and middle: 	(A) section of the NEPOMUC beamline, 
		(B) first beam monitor (BM\,1) with optional apertures and electrostatic accelerator, 
		(C) brightness enhancement unit with BM\,2 and first electrostatic lens system for beam focusing onto the remoderation foil,
		(D1) second electrostatic lens system for acceleration of the re-emitted slow positrons and beam focusing onto the sample, 
		(D2) sample chamber with optional BM\,3 and piezo positioning system for sample scanning.
		The HPGe-detectors are highlighted (green) and gray dotted lines mark the planes for the beam shape measurements (BM\,1-3).
		Right: Simulation of positron trajectories focused onto the remoderation foil
		}
		\label{CDB_Spectrometer}
\end{figure}

The magnetic and electrostatic beam guidance as well as the brightness enhancement device are designed in such a way to allow the use of both, the primary and the remoderated beam of NEPOMUC.
As shown in figure\,\ref{CDB_Spectrometer} the upgrade of the CDB spectrometer comprises the beam guidance realized by electrostatic lenses and magnetic field coils, the brightness enhancement system, beam monitors, an electrostatic accelerator, and the new sample chamber with sample positioning system. 
Based on first calculations of the beam transport \cite{Tgigl2014} more detailed simulations of the positron trajectories are depicted in figure\,\ref{CDB_Spectrometer} as well.

The positron beam of the NEPOMUC beamline enters the spectrometer via a bend into the first section of the spectrometer which basically consists of the electrostatic accelerator and the focusing unit of  the brightness enhancement system (see A-C in figure\,\ref{CDB_Spectrometer}).
For practical reasons, the coils for the longitudinal magnetic guiding  field are installed in a Helmholtz-like geometry.  
The $\mu$-metal shielding minimizes the influence of external stray fields and additional saddle coils for transverse magnetic fields can be used to optimize the beam guidance on-axis. 
Due to its superior importance, brightness enhancement as well as details of the remoderation system and its performance are explained in detail in section \ref{sec:be}. 

Three new beam monitor units (BMs) are installed in order to determine the shape and the intensity of the beam.  The first two BMs consist of a stack of micro channel plates (MCP) with an additional phosphor scintillation screen and a CCD camera \cite{NucInstr_mcp,Hug08d}.
At the position of BM\,1 apertures of different size can be inserted for further reduction of the beam diameter.
The energy dependent beam spot at the sample position can be directly monitored by using a blank phosphor screen (BM\,3) without a MCP, i.e.\,without additional electric fields which might influence the beam shape. 
The resolution limit is given by imaging of the phosphor screen with the CCD camera (Basler scout scA1390-17gm, pixel size: 4.65\,$\mu$m$\times$4.65\,$\mu$m) and an optic with 35\,mm focal length resulting in 23.25\,$\mu$m per pixel. 
Hence, the visual feedback reduces dramatically the time for tuning the magnetic and electrostatic fields in order to optimize the beam focus at various implantation energies.

After passing the remoderator, an acceleration system consisting of eight electrodes is used to adjust the energy of the positron beam for measurements with either the NEPOMUC beam (without additional remoderation) or the threefold moderated beam.
Finally, the second electrostatic lens system is used to focus the beam onto the sample in the analysis chamber (see D1,2 in figure\,\ref{CDB_Spectrometer}).
For depth dependent measurements the sample can be biased up to -30\,kV enabling a positron implantation of up to a few $\mu$m. 
A 2D piezo positioning system with optical encoders allows to scan an area of 19$\times$19\,mm$^2$ with high accuracy and repeatability in the nm-range.
Due to the geometry of the sample holder the maximum height of the sample is limited to 10\,mm. 

The design of the new sample chamber allows for a quick setup change for measurements at high and low temperatures. 
Temperature dependent in situ (C)DBS can be performed from room temperature up to 1000\,K by replacing the high resolution position unit with a heatable sample holder \cite{Rei13}.
The light of a 250\,W halogen lamp, which is located in one of the focuses of an ellipsoidal Cu reflector, is concentrated onto the back side of the sample in the opposite focus. 
The reflector is pressed in a water cooled cup of non-magnetic steel \cite{Rei13}. 
For realizing positron implantation energies of up to 30\,keV, the sample holder is electrically insulated by three Macor pins separating the sample holder ring from the reflector. 
The sample is located in the middle of a ``spider`` holder consisting of five Mo cantilevers fixed on the Macor insulators to minimize the heat loss by heat conduction. 
Low temperature experiments down to 40\,K can be accessed by a closed-cycle He cryostat. 
The Cu sample holder is mounted on a cold head via a sapphire insulator for thermal coupling and electrical insulation. 
Despite the large devices for heating and cooling, respectively, the sample can still be moved within a reduced scanning range of about 5$\times$5\,mm$^2$.

The sample chamber can be surrounded by up to eight high purity Ge detectors which are conventionally used in (C)DBS. 
The detectors are slightly inclined with respect to the sample plain in order to reduce loss in count rate due to absorption or scattering of the annihilation radiation in the sample itself and structural material of the sample chamber.
At present, four Ge detectors (30\% efficiency, energy resolution of 1.3\,keV at 477.6\,keV \cite{Sta07b}) are used in routine operation either separately or in coincidence mode for DBS and CDBS, respectively. 

%++++++++++++++++++++++++++++++++++++++++++++++++++++++++++++++++++++++++++++++++++
\subsection{Brightness enhancement system}
\label{sec:be}
%++++++++++++++++++++++++++++++++++++++++++++++++++++++++++++++++++++++++++++++++++

The smallest achievable beam diameter is limited by the (transverse) phase space volume occupied by the ensemble of particles. 
This limit is a consequence of Liouville's theorem, which states, that the phase space density remains constant under the influence of conservative forces. 
For this reason, the source properties are of highest importance since they define the occupied phase space volume and hence the final limit of the beam diameter.
At a given intensity $I$ and kinetic energy in forward direction $E_\parallel$ (sometimes called \textit{longitudinal} energy), the brightness $B$ of the beam can be expressed in terms of its diameter $d$ and divergence $\theta$ \cite{Coleman2000,Mil80c,Sch88,Hug12a}:

\begin{equation}
B = \frac{I}{d^2 \theta^2 E_\parallel}= \frac{I}{d^{2}p_\perp^{2}/2m_e}
\end{equation}

In order to circumvent this limit, i.e. to further reduce  the transverse momentum $p_\perp$, non-conservative forces are used to cool the positrons by interaction with matter. 
During this so called (re)moderation process, positrons are implanted in a material exhibiting a negative positron work function such as W, Pt or Ni. 
After thermalization and diffusion to the surface, low-energy (moderated) positrons are reemitted predominantly perpendicular to the surface with a kinetic energy defined by the absolute value of the (negative) positron work function $\Phi^+$ ($\Phi^+_{\mathrm{W}}$\,=\,$-3.0$\,eV \cite{Coleman2000}, $\Phi^+_{\mathrm{Pt}}$\,=\,$-1.95$\,eV \cite{Hug02d} or $\Phi^+_{\mathrm{Ni}}$\,=\,$-1.4$\,eV \cite{Sch86}). 
In order to achieve a high moderation efficiency the applied poly- or single- crystalline moderators should exhibit a low defect concentration leading to long positron diffusion length as well as clean and flat surfaces.
In principle, high brightness positron beams can be achieved by multiple moderation, i.e. by repeating acceleration, focusing and moderation in several \textit{remoderation} stages \cite{Mil80c}.
Details of positron  (re)moderation can be found elsewhere (see e.g. \cite{Hug16} and references therein, and \cite{Coleman2000} for various moderator geometries). 

For the realization of the positron microbeam in the CDB spectrometer upgrade  the positrons are moderated three times: 
First, positrons are created by pair production and moderated (so-called \textit{selfmoderation}) in annealed polycrystalline Pt foils inside the positron source  NEPOMUC \cite{Hug08b}.
Then, the brightness of the primary 1\,keV positron beam with 10$^9$ moderated positrons per second \cite{Hug14b} is enhanced by a W(100) single crystal remoderator operated in back reflection geometry \cite{Pio08}. 
Compared to a transmission geometry the beam guidance is much more sophisticated but thicker W crystals can be used that facilitates heating and long term operation. 
For most experiments the energy of the remoderated beam is set to 20\,eV and the beam diameter amounts to $<$\,2\,mm (FWHM) in a 7\,mT guiding field \cite{NEPupg,Pio08}.
Finally, in the new setup of the CDB spectrometer we apply a Ni(100) foil acting as transmission remoderator to generate the positron microbeam.

In contrast to the reflexion geometry the beam optics becomes much simpler for transmission remoderators due to the rotational symmetry of all optical components.
However, in transmission geometry thin foil moderators have to be applied in order to achieve a high moderation efficiency.
In our setup we use a free-standing single crystalline Ni(100) foil with a thickness of 100\,nm which results in an optimum implantation energy in the order of 5\,keV for a maximum yield of remoderated positrons. 
Compared to W, Ni has several advantages: easier preparation of thin foils, lower density, significantly lower annealing  temperature, and the energy distribution of the remoderated positrons was shown to be more narrow resulting in a smaller beam spot after focusing \cite{Sch83,Sch86}. 

At the CDB spectrometer an electrostatic accelerator in the longitudinal magnetic guiding field (figure\,\ref{CDB_Spectrometer}\,B) allows the adjustment of the positron implantation energy.
Especially focusing the beam onto the remoderator foil either by a magnetic or an electrostatic lens system is crucial. 
In a similar system, Oshima et al.\,\cite{Osh08} focus the beam by a magnetic lens system which had to be optimized for the suppression of the magnetic field at the position of the moderator and hence to avoid the introduction of transverse momentum to the moderated positrons. 
Another disadvantage is the large extent of the magnetic lens including the magnetic yokes for guiding the magnetic field to the focusing region.
In order to circumvent these difficulties we designed a very compact electrostatic focusing unit consisting of three electrodes as shown in figure\,\ref{CDB_Spectrometer}\,C. 
A magnetic field termination made of metallic glass stripes on a $\mu$-metal support is mounted in front of the focusing lenses of the brightness enhancement system.
Hence, disturbances by magnetic fields of the purely electrostatic guiding and focusing system, and, more importantly, residual magnetic fields affecting the trajectories of the slow positrons reemitted from the remoderator surface can be avoided. 
Simulation of the positron trajectories have been performed using the COMSOL multiphysics package to optimize the new layout of the optical components. 
As shown in figure\,\ref{CDB_Spectrometer}\,(right) an incoming beam of 2.5\,mm diameter can be electrostatically focused to a spot of 0.5\,mm diameter on the remoderation foil and extracted on the backside.
The re-emitted slow positrons are accelerated, formed to a beam and focused onto the sample by a second electrostatic lens system (figure\,\ref{CDB_Spectrometer}\,D1,2).

The beam spot at the position of the remoderation foil can be monitored by an insertable beam monitor (BM\,2) consisting of a MCP module similar to BM\,1. 
This enables the inspection of the beam spot and a quick tuning of the parameters of the first electrostatic lens system for optimizing the focus on the remoderator foil.  
Optional apertures can be mounted on top of the remoderation foil for further reduction of the beam spot.

%++++++++++++++++++++++++++++++++++++++++++++++++++++++++++++++++++++++++++++++++++
\subsection{Preparation and characterization of the Ni(100) remoderation foil}
%++++++++++++++++++++++++++++++++++++++++++++++++++++++++++++++++++++++++++++++++++

Conditioning of the Ni(100) transmission remoderator significantly increases the yield of  reemitted positrons \cite{Fuj08,Osh08}.
The preparation of the remoderation foil comprises annealing in vacuum as well as additional oxygen and hydrogen treatment for removing surface impurities.
In order to get detailed information on the conditioning, the remoderation foil was characterized by x-ray induced photo electron spectroscopy (XPS) and temperature dependent DBS. 
Furthermore,  the optimum positron implantation energy was determined using the former CDB spectrometer \cite{Rei13}. 

\begin{figure}[htb]
	\centering
	\includegraphics[width=9cm]{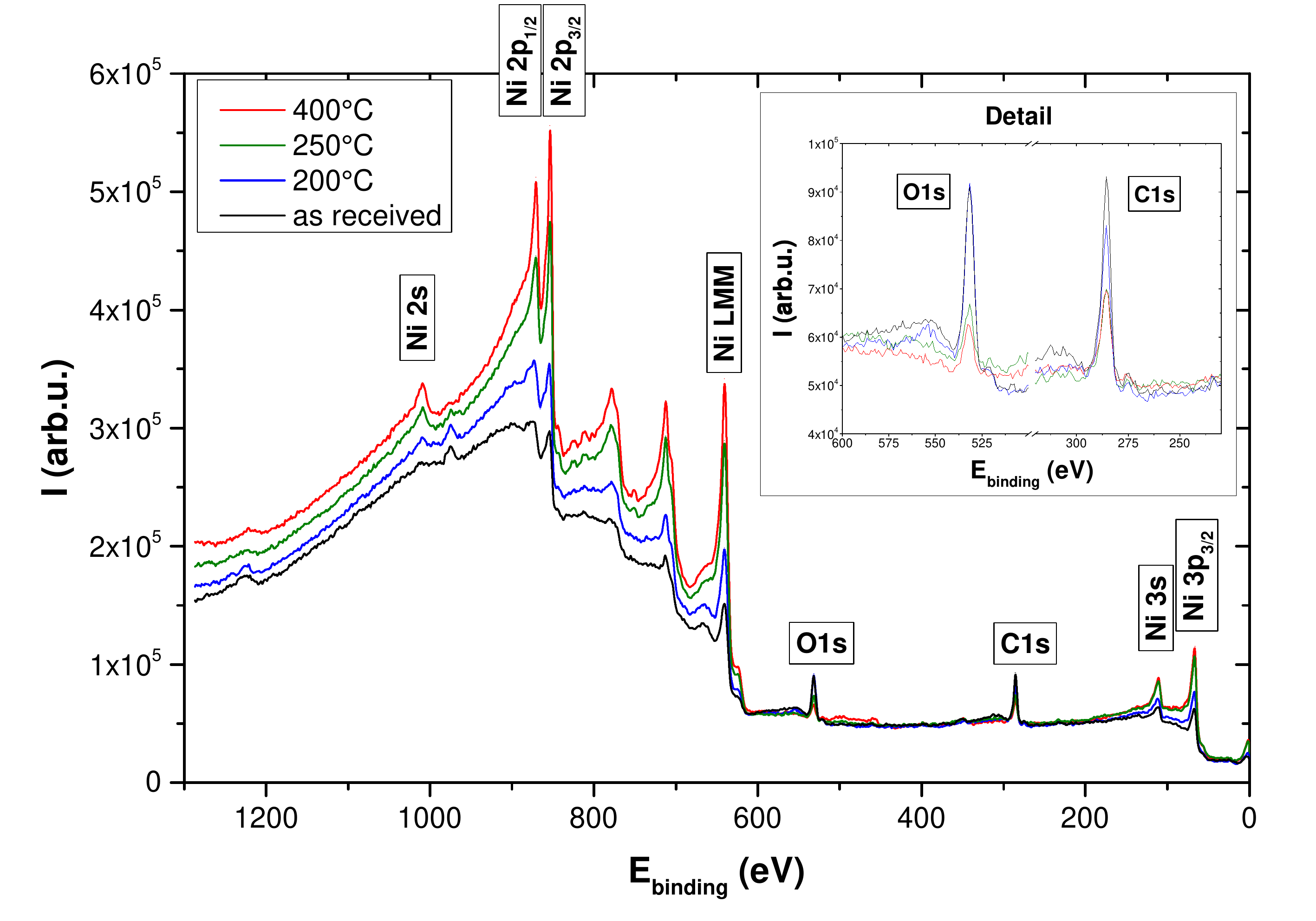}
	\caption{XPS-spectra of Ni(100) at various temperatures: Heating the Ni(100) foil to 400\,$^\circ$C leads to an increase of the Ni signal by a factor of six whereas the O and C signal is significantly reduced (insert).}
	\label{XPS_Ni}
\end{figure}

The removal of the surface impurities C and O could be clearly observed by temperature dependent XPS.
As shown in the XPS spectra (figure\,\ref{XPS_Ni}), heating the foil from room temperature to 400\,$^\circ$C causes an increase in the Ni signal by a factor of six. Simultaneously, the C and O peaks decrease by factors of two and four, respectively (see insert in figure\,\ref{XPS_Ni}).
Hence, just by heating the surface, contaminations can be significantly reduced resulting in a higher remoderation yield. 
Fujinami et al.\,showed that heat treatment in an O and H atmosphere lead to a further reduction of surface impurities \cite{Fuj08}.

\begin{figure}[htb]
	\centering
  \subfigure[]{\includegraphics[width=0.48\textwidth]{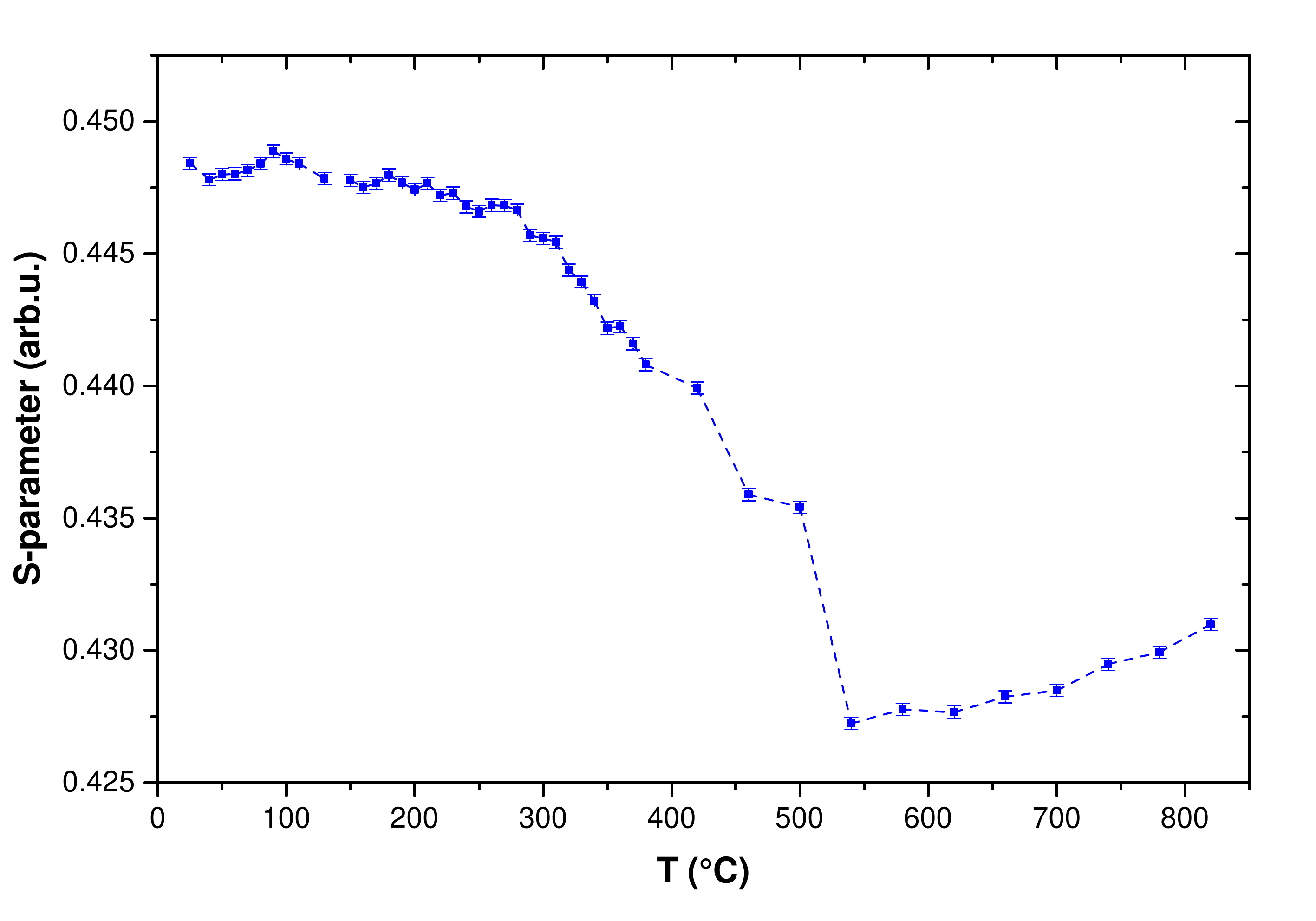}}
	\subfigure[]{\includegraphics[width=0.48\textwidth]{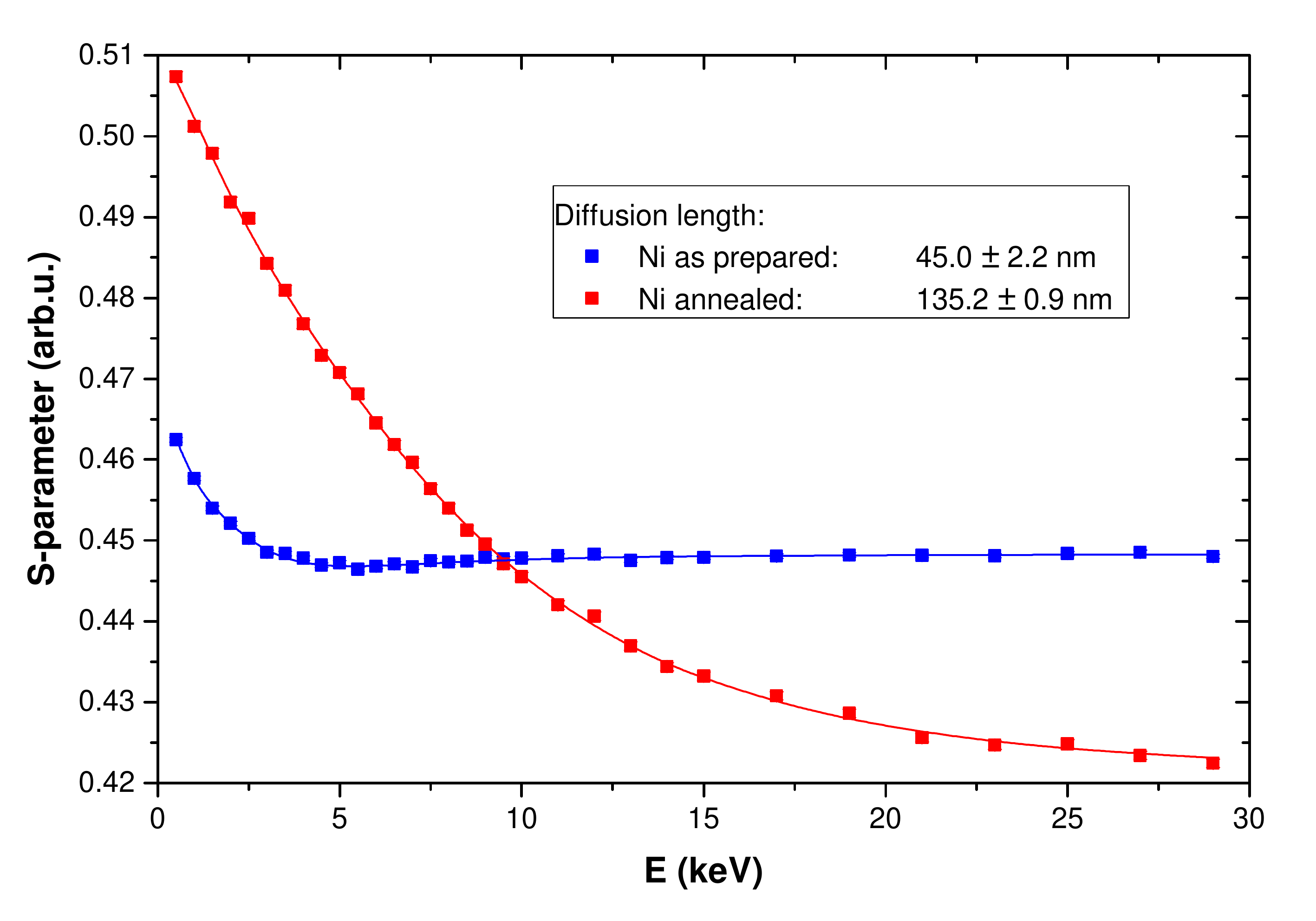}}
	\caption{(a) Temperature dependent DBS on polycrystalline Ni: The minimum S-parameter at 540\,$^\circ$C indicates fully annealing. At higher temperatures the S-parameter rises due to the thermal expansion of the crystal lattice and thermally created vacancies.
(b) Depth profiles of the S-parameter: The positron diffusion length before and after tempering is determined by fitting the data with VEPFIT \cite{vVe91} (solid lines).}
	\label{Ni_annealing_CDB}
\end{figure}

The annealing behavior of the new Ni(100) foil was studied by temperature dependent DBS at a positron implantation energy of 30\,keV. 
Figure\,\ref{Ni_annealing_CDB} shows the bulk S-parameter, which is considered to be related to the concentration of open-volume defects in the sample, as function of temperature of the Ni foil.  
A significant drop in the S-parameter can be observed between 300\,$^\circ$C and 550\,$^\circ$C indicating the diffusion of vacancies and hence annealing of the Ni crystal.
The found annealing temperature of about 540\,$^\circ$C, i.e. minimum value of S, is in good agreement with the literature \cite{BoerNi}.
By further increasing the temperature to about 820\,$^\circ$C the S-parameter rises again due to the thermal expansion of the crystal lattice and creation of vacancies.

In order to obtain the highest moderation efficiency the thermalized positrons should reach the (back) surface of the Ni foil with high probability. 
Therefore, positron trapping in lattice defects should be minimal to achieve a large positron diffusion length. 
Values of the positron diffusion length are obtained from depth dependent measurements of the S-parameter with DBS of a (polycrystalline)  Ni sample at room temperature as well as after the heat treatment. 
As shown in figure\,\ref{Ni_annealing_CDB}\,b the S-parameter for the as-received sample reaches its bulk value at about 5\,keV implantation energy whereas after tempering the S-parameter saturates at about 25\,keV. 
The S value in the bulk of the annealed Ni is significantly reduced compared to the as-prepared state as expected due to its lower defect concentration.
The results for the diffusion length obtained by VEPFIT \cite{vVe91} are 45.0\,nm and 135.2\,nm for the polished as-prepared and the annealed sample, respectively. 
Consequently, the single crystalline Ni foil as used in the upgraded CDB spectrometer is well suitable as moderator since the positron diffusion length is in the order of, and even larger than, the foil thickness of 100\,nm. 

\begin{figure}[htb]
	\centering
	\includegraphics[width=9cm]{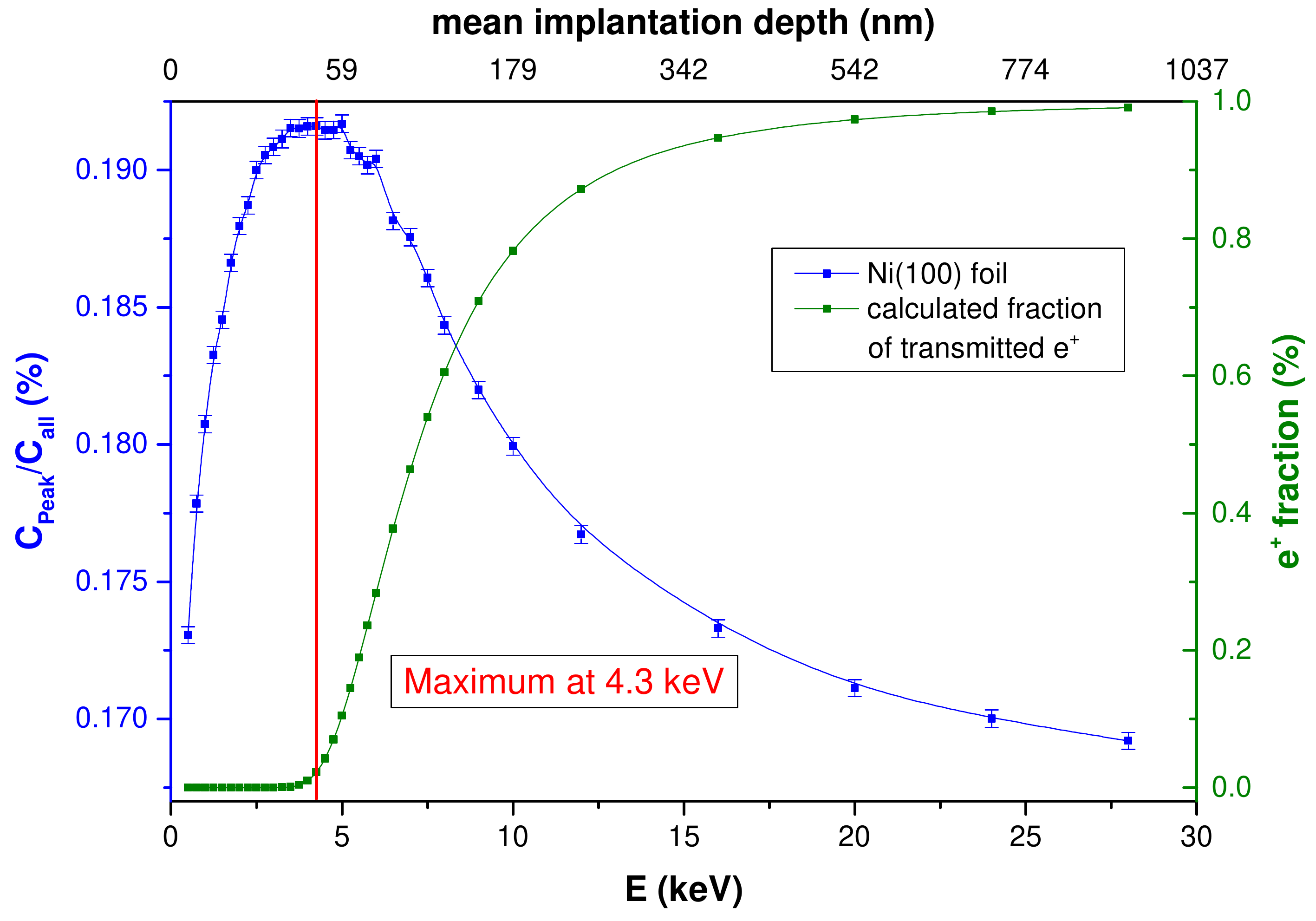}
	\caption{Fraction of positrons annihilating in the 100\,nm thick Ni(100) foil at different implantation energies. 
	The fraction of positrons annihilating inside the remoderation foil is determined from the amount of counts in the 511\,keV photopeak $c_{peak}$ divided by the total counts $c_{all}$ of the  spectrum. 
	Most of the positrons are implanted in the foil at an energy of 4.3\,keV according to a mean implantation depth of about 45\,nm. 
	Only 2.2\,\% of the positrons pass the foil as calculated by the Makhovian implantation profile (right axis).}
	\label{Ni_C_peak_C_total}
\end{figure}

In order to maximize the amount of remoderated positrons emitted from the Ni(100) foil the energy of the positrons has to be low enough that they are fully thermalized but sufficiently high that a maximum fraction reaches the backside of the foil.
Therefore, we determined the appropriate positron implantation energy by measuring the energy dependent annihilation rate in the remoderation foil.
Figure\,\ref{Ni_C_peak_C_total} shows the fraction of the measured counts in the 511\,keV photopeak with respect to the total counts $c_{peak}/c_{all}$ as a function of positron implantation energy, which serves as indicator for the fraction of positrons annihilating inside the remoderation foil. 
The amount of annihilating positrons clearly increases with higher positron energy until a maximum is reached at around 4.3\,keV. 
At higher energy the positrons pass the foil without being (fully) moderated resulting in a decreased annihilation rate in the Ni(100) foil. 
The implantation energy of 4.3\,keV, according to a mean implantation depth of about 45\,nm, obviously ensures that a maximum amount of positrons can reach the backside of the foil within the diffusion length of 135\,nm. 
The fraction of not (fully) remoderated transmitted positrons passing through the foil was calculated to be 2.2\,\%. 

%++++++++++++++++++++++++++++++++++++++++++++++++++++++++++++++++++++++++++++++++++
\section{Performance of the positron microbeam}
%++++++++++++++++++++++++++++++++++++++++++++++++++++++++++++++++++++++++++++++++++

The CDB spectrometer is upgraded to handle both beams provided by  the positron source NEPOMUC \cite{NEPupg}, the primary beam with an energy of 1\,keV and a diameter of 7\,mm (FWHM) and the remoderated 20\,eV beam with a diameter of 2\,mm  \cite{Hug12b,NEPupg}.  
As in most experiments, the remoderated 20\,eV beam is used for all measurements presented here. 
The beam is magnetically guided via a beam switch  to the CDB spectrometer where it can be used directly or further enhanced in brightness by the Ni(100) transmission remoderator. 
Note that finally a three-fold moderation is used to generate the positron microbeam.
In the following, the beam without further remoderation is called "NEPOMUC remoderated beam" and the three-fold moderated beam with the Ni(100) transmission remoderator of the spectrometer is called "positron microbeam".

\begin{figure}
\centering
	\subfigure[]{\includegraphics[width=0.3\textwidth]{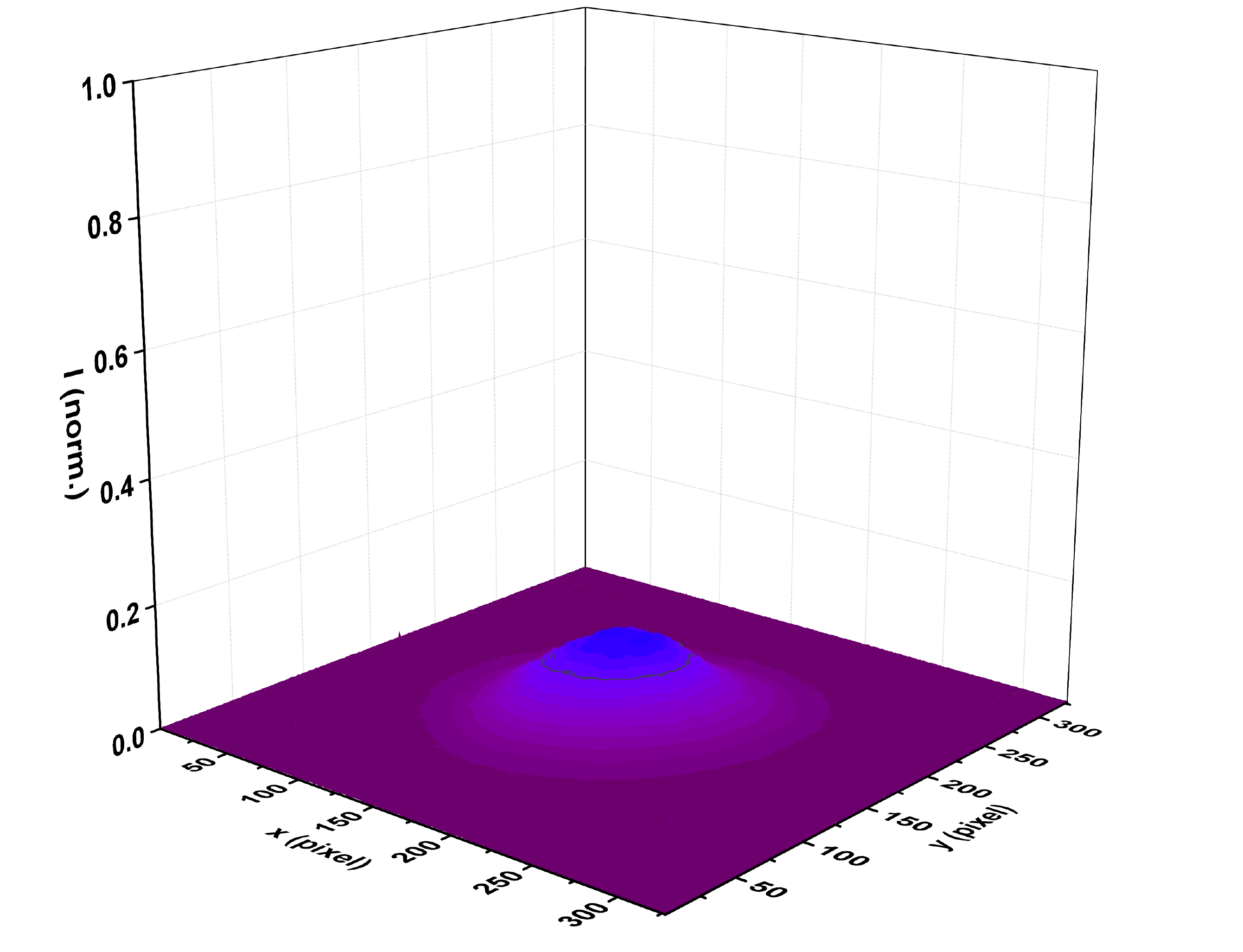}}
	\subfigure[]{\includegraphics[width=0.3\textwidth]{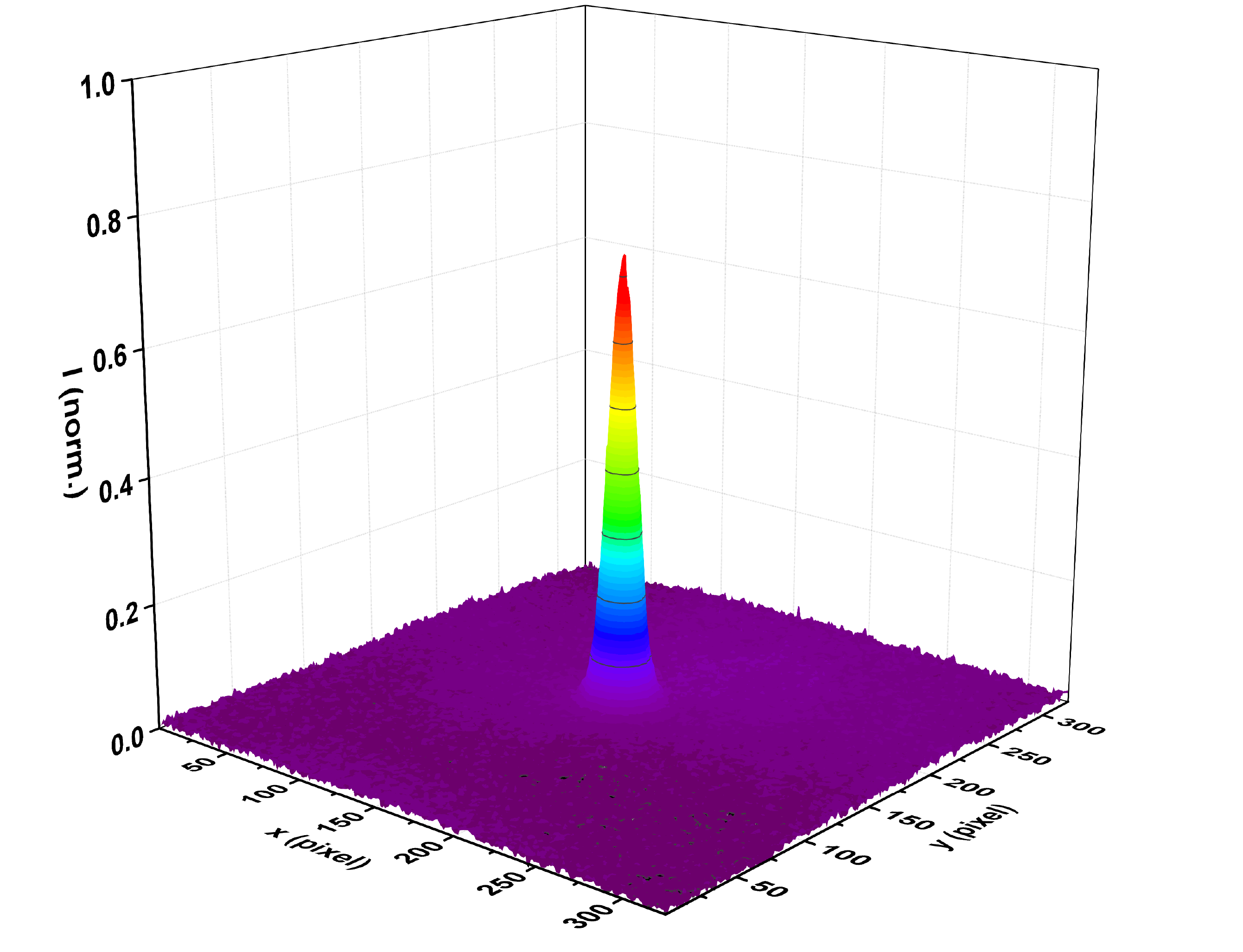}}
	\subfigure[]{\includegraphics[width=0.3\textwidth]{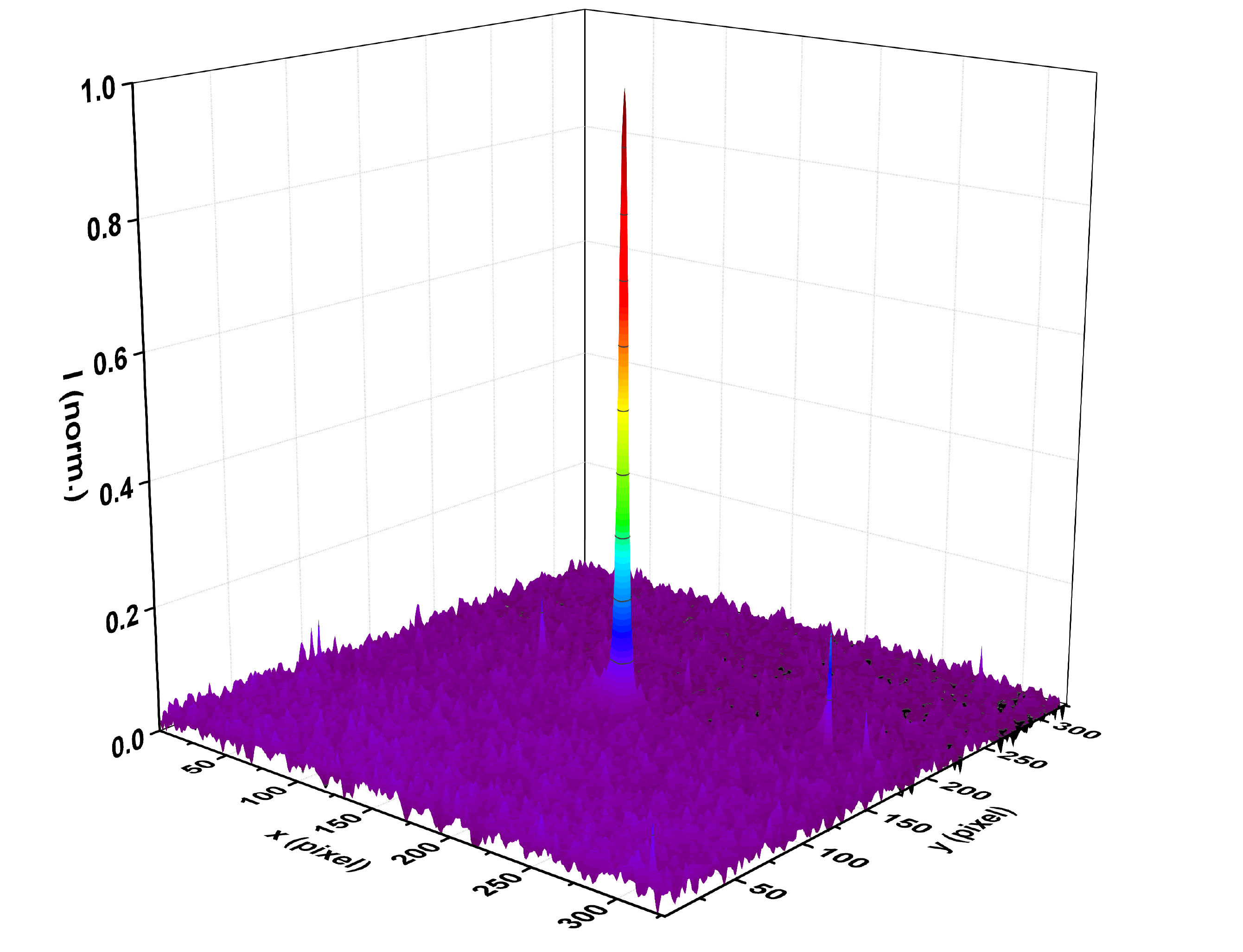}}
	\caption{3D plots of the intensity profiles of the NEPOMUC remoderated beam without Ni remoderation foil at the three beam monitors (BM): 
	(a) 20\,eV beam at BM\,1 as provided by the NEPOMUC beamline with a diameter of about 2.5\,mm (FWHM), 
	(b) beam accelerated and focused by the first electrostatic lens system onto BM\,2 at the position of the remoderation foil (d$_{FWHM}\approx$\;0.5\,mm), and 
	(c) beam profile on the phosphor screen (BM\,3) at the sample position with a diameter of 0.25\,mm (FWHM) and an energy of 30\,keV.
	As transport loss is negligible the intensity profiles are normalized to same integrated counts. 
	}
	\label{Beamprofiles}
\end{figure}

The performance of the various acceleration and focusing units of the spectrometer is tested by comparing the beam profiles detected with various beam monitors (compare figure\,\ref{CDB_Spectrometer}).
The positron intensity profiles shown in figure\,\ref{Beamprofiles} are normalized to the same value since no apertures have been used which might have lead to transport loss in the electrostatic beam guiding system.
The 20\,eV beam as provided by NEPOMUC enters the spectrometer with a diameter of about 2.5\,mm at BM\,1 (figure\,\ref{Beamprofiles}\,a).
After passing through the accelerator and the first focusing system of the brightness enhancement unit, a beam spot of 0.5\,mm in diameter is achieved (figure\,\ref{Beamprofiles}\,b) and the beam energy is set to 5\,keV for positron remoderation in the Ni(100) foil. 
As shown in figure\,\ref{Beamprofiles}\,c, even without the remoderation foil the diameter of the NEPOMUC remoderated beam can be reduced to 0.25\,mm at the sample position.  
In contrast to the previous setup of the CDB spectrometer \cite{Rei15a}, the beam diameter at the maximum implantation energy of 30\,keV could be reduced by 16\,\%.
This improvement is mainly attributed to the more homogeneous magnetic beam guidance with $\mu$-metal shielding, the use of non-magnetic materials for building the new instrument and the new designed electrostatic lens system for beam focusing. 

\begin{figure}
	\centering
	\includegraphics[width=0.8\textwidth]{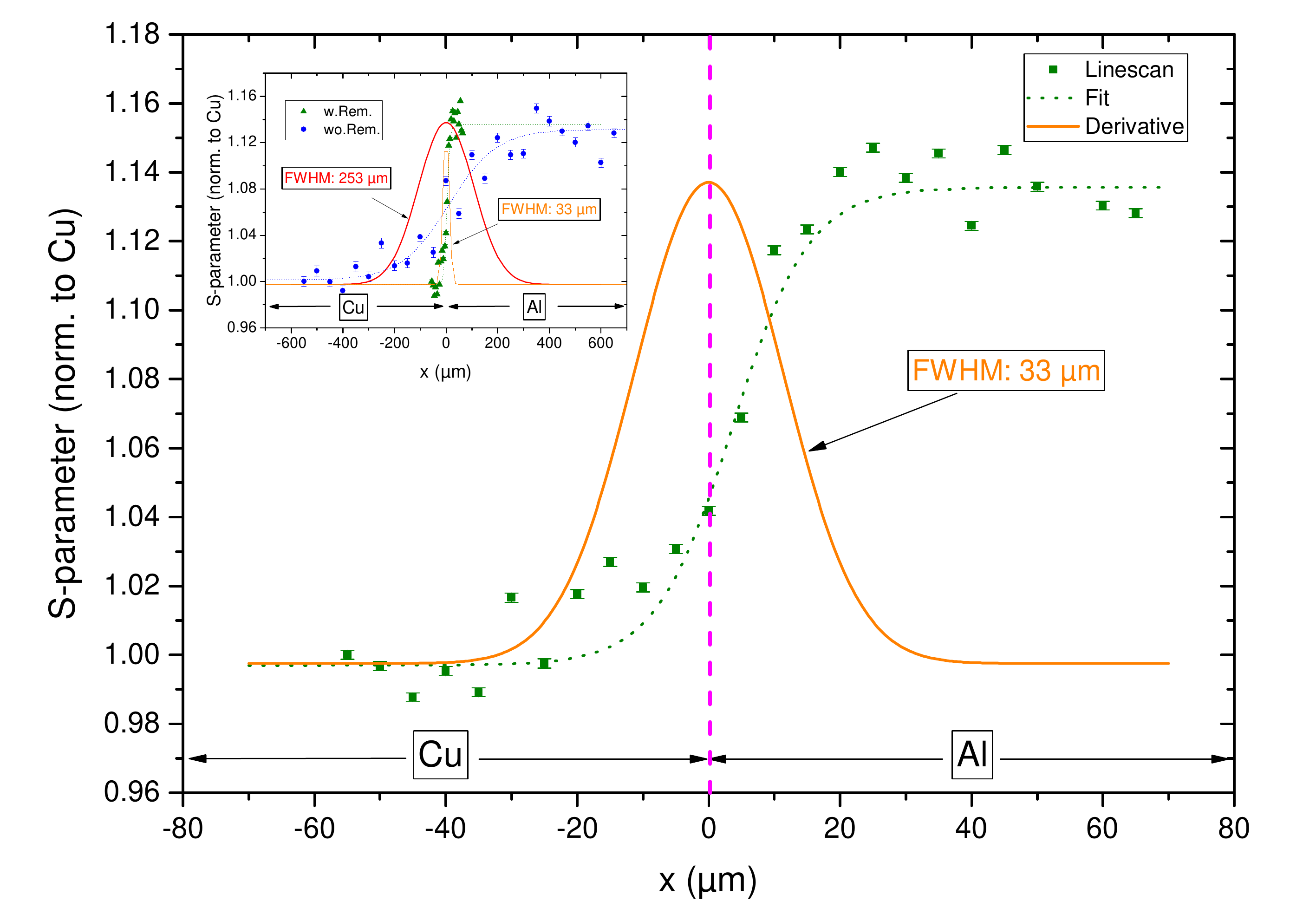}
	\caption{Diameter of the positron microbeam and the NEPOMUC remoderated beam (insert) determined by DB line scans over a sharp Cu/Al edge at maximum beam energy. 
			A beam diameter of 33(7)\,$\mu$m is obtained for  the positron microbeam and 253(40)\,$\mu$m for the NEPOMUC remoderated beam without additional remoderation by fitting the data with an error function (dashed lines, derivative plotted as solid lines).
	}
	\label{Linescanvergleich}
\end{figure}

Due to the limited resolution of BM\,3, the diameter of the positron microbeam at the sample position was determined using the so-called knife edge technique \cite{Hug02d}.
For this purpose, a special prepared sample with an Cu/Al edge is used as these two materials provide a distinct contrast of the measured S-parameter.  
Two line scans across the Cu/Al edge were performed with the positron microbeam (green curve) and the NEPOMUC remoderated beam (blue curve) as shown in figure\,\ref{Linescanvergleich}. 
As expected, the position dependent S-parameter S(x) shows a clear transition  from low to high at the Cu/Al edge. 
An error function was fitted to the data (dashed lines) in order to determine the beam diameter (FWHM) from its derivative (solid lines in figure\,\ref{Linescanvergleich}).
Thus a diameter of 253\,$\mu$m\,$\pm$\,40\,$\mu$m at a beam energy of 30\,keV was measured for the NEPOMUC moderated beam in agreement with the beam profile shown in figure\,\ref{Beamprofiles}\,c.
Using the transmission remoderator at a beam energy of 25\,keV leads to a diameter of 51\,$\mu$m\,$\pm$\,11\,$\mu$m. 
With an additional aperture in front of the remoderation foil the diameter of the positron microbeam could be further reduced to 33\,$\mu$m\,$\pm$\,7\,$\mu$m.

\begin{figure}[ht]
\centering
\begin{minipage}{0.28\textwidth}
\includegraphics[width=0.9\textwidth]{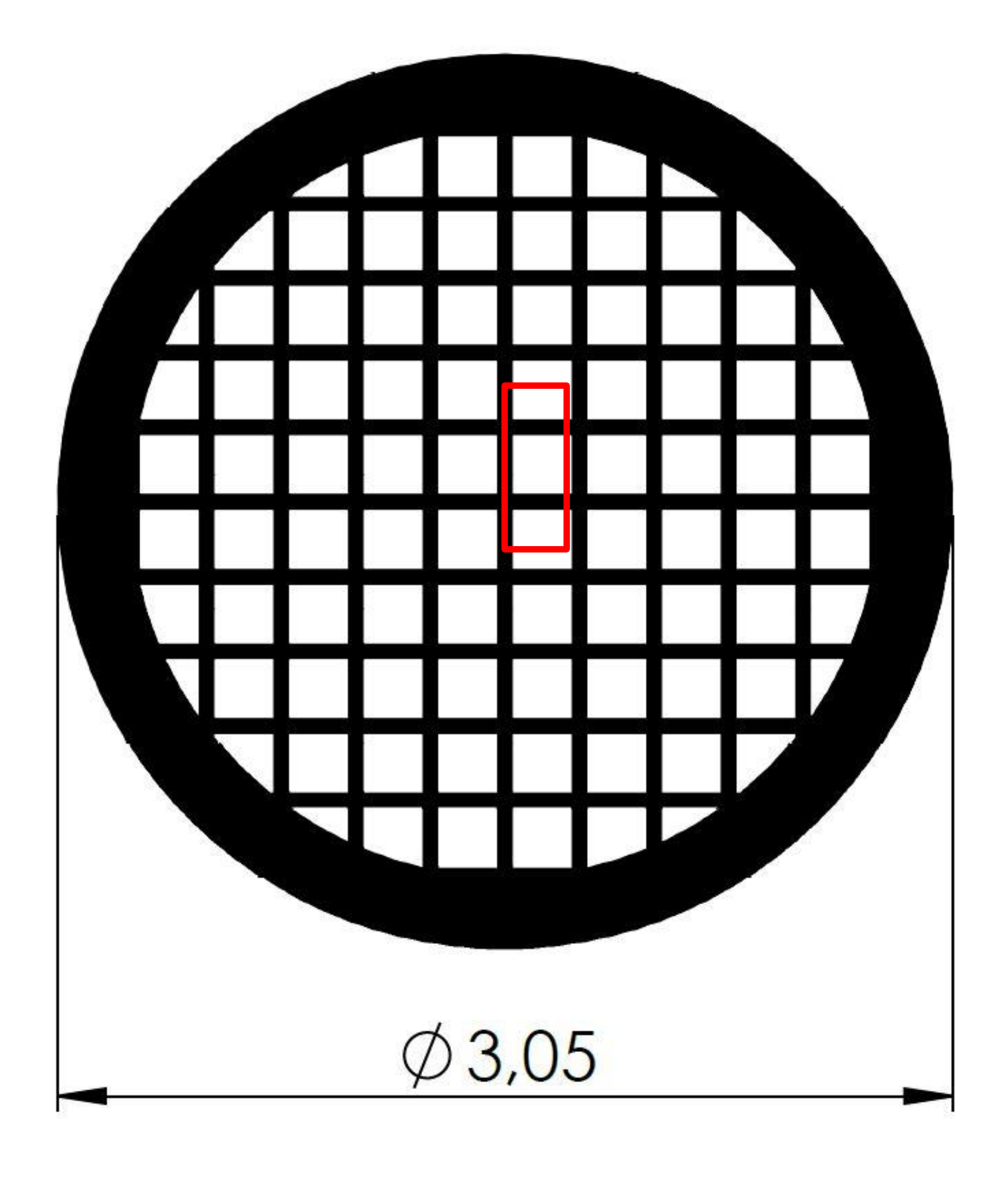}
\end{minipage}
\begin{minipage}{0.28\textwidth}
\includegraphics[width=1\textwidth]{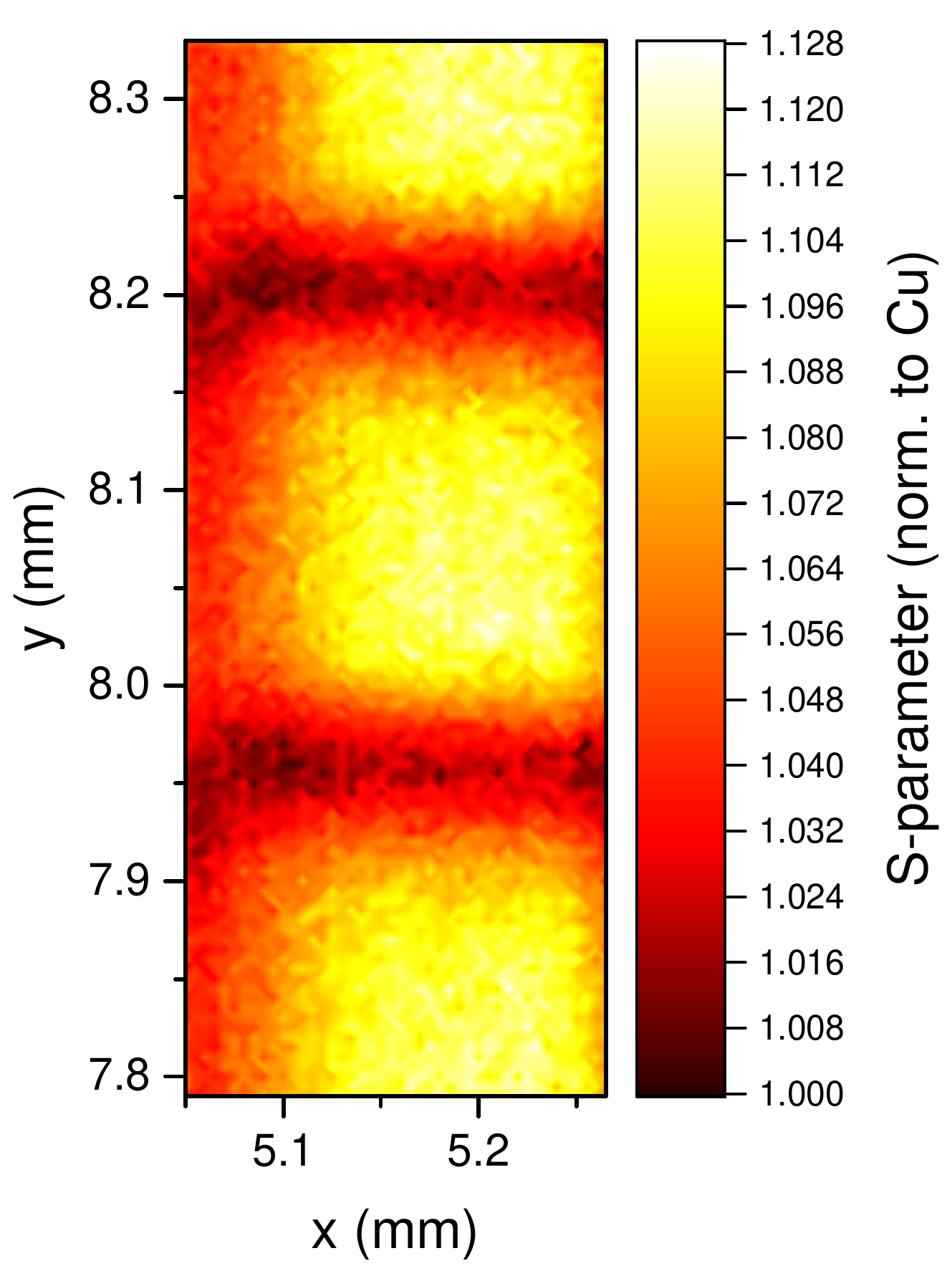}
\end{minipage}\\
\subfigure[Cu mesh]{\begin{minipage}[b]{0.18\textwidth}~\end{minipage}}
\subfigure[S-parameter map]{\begin{minipage}[b]{0.28\textwidth}~\end{minipage}}
\caption{2D S-parameter map of a Cu mesh.
(a) Cu mesh with 50\,$\mu$m thick bars and 204\,$\mu$m spacing mounted on Al. The scan area of the positron beam is marked (red rectangle). 
(b) S-parameter map recorded with the positron microbeam without aperture ($\Delta_{x,y}$\,=\,5\,$\mu$m, t$_m$\,=\,7\,s). Note the distortion-free imaging of the mesh.}
\label{Mesh100}
\end{figure}

\begin{figure}
	\centering
    \begin{minipage}{\textwidth}
	\subfigure[]{\includegraphics[width=0.21\textwidth]{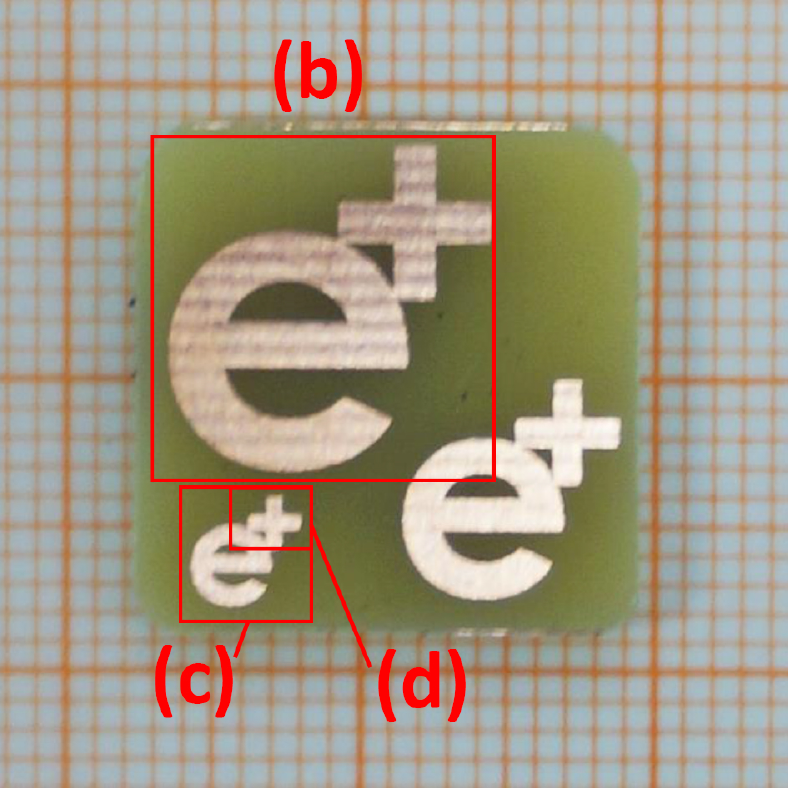}}
	\subfigure[]{\includegraphics[height=36mm]{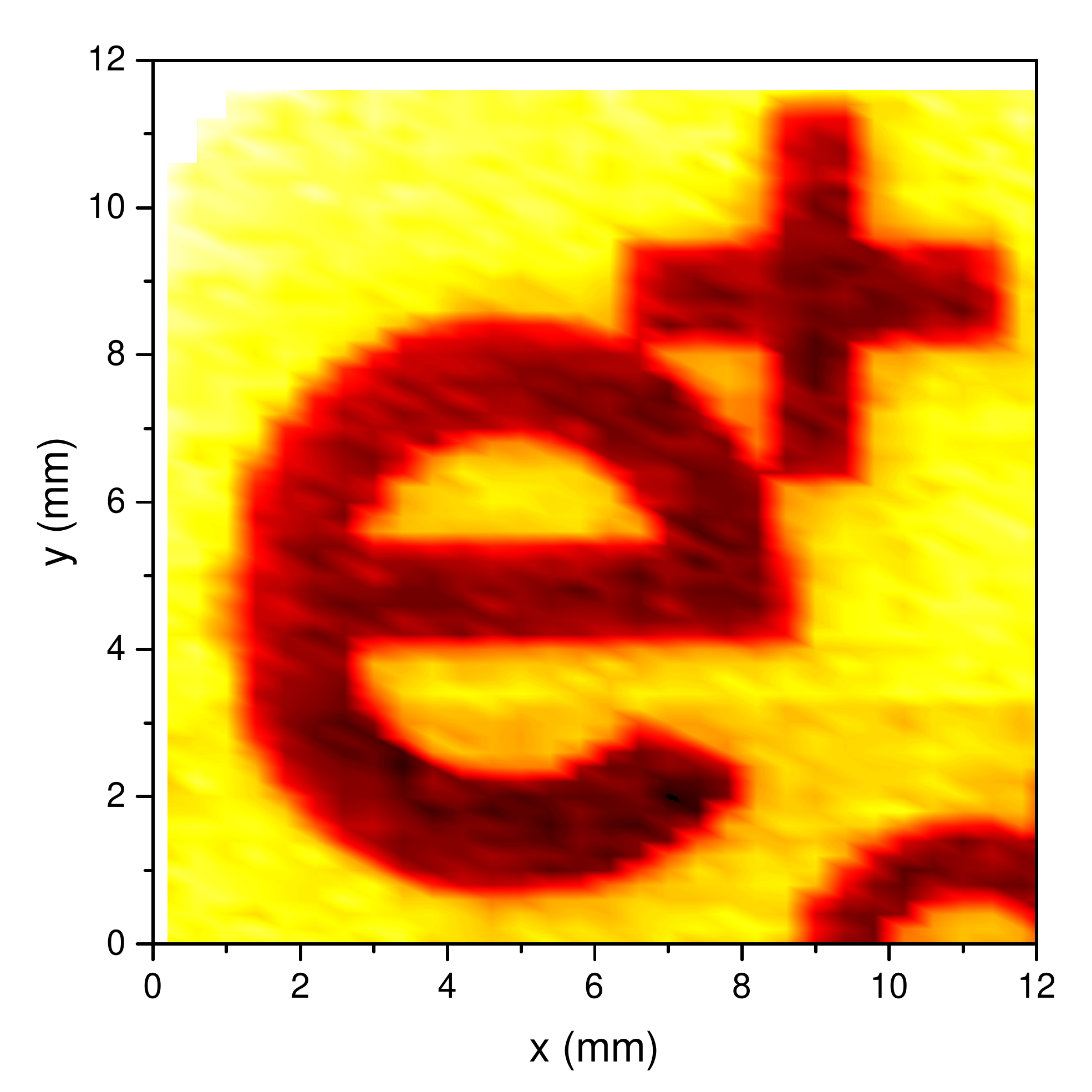}}
	\subfigure[]{\includegraphics[height=36mm]{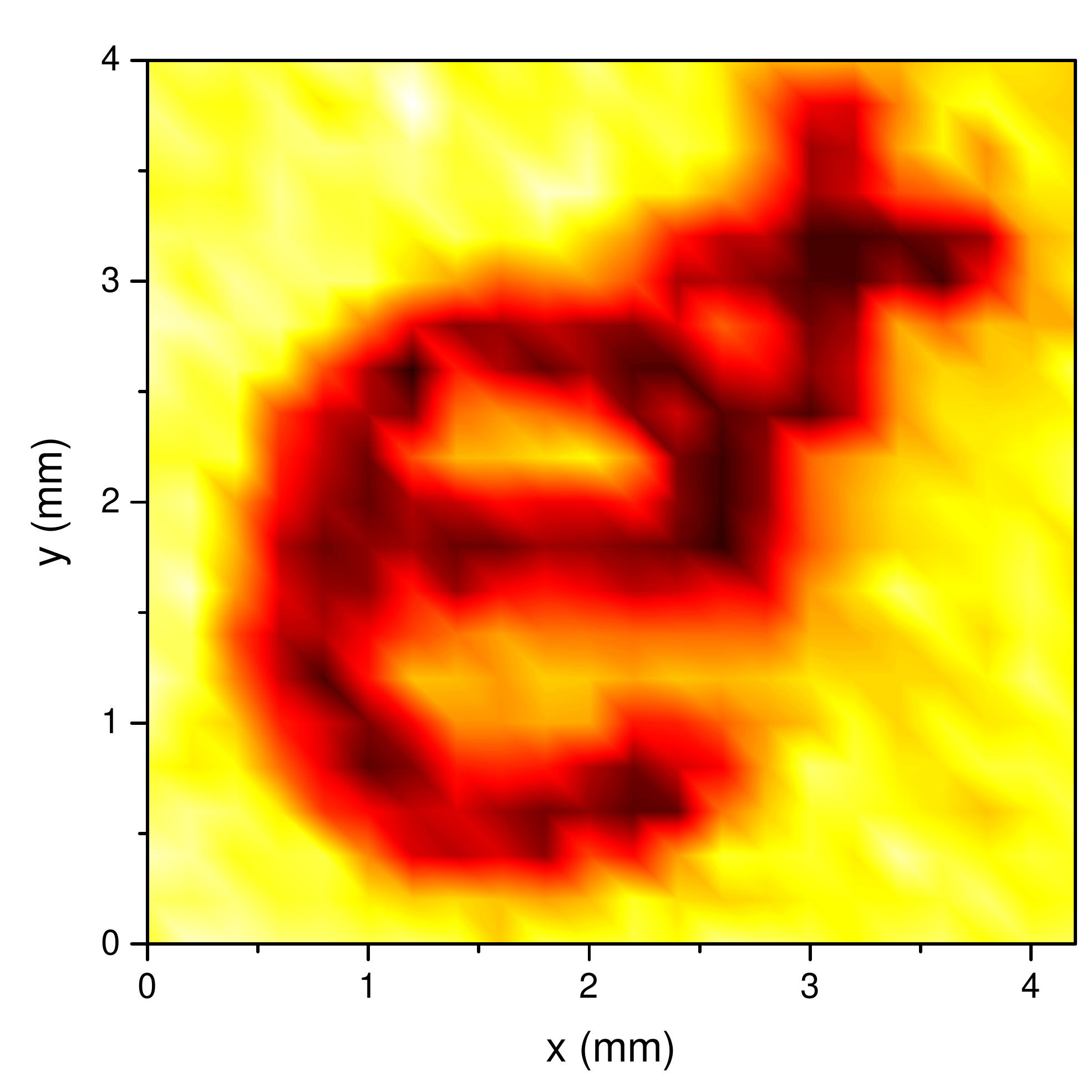}}
	\subfigure[]{\includegraphics[height=36mm]{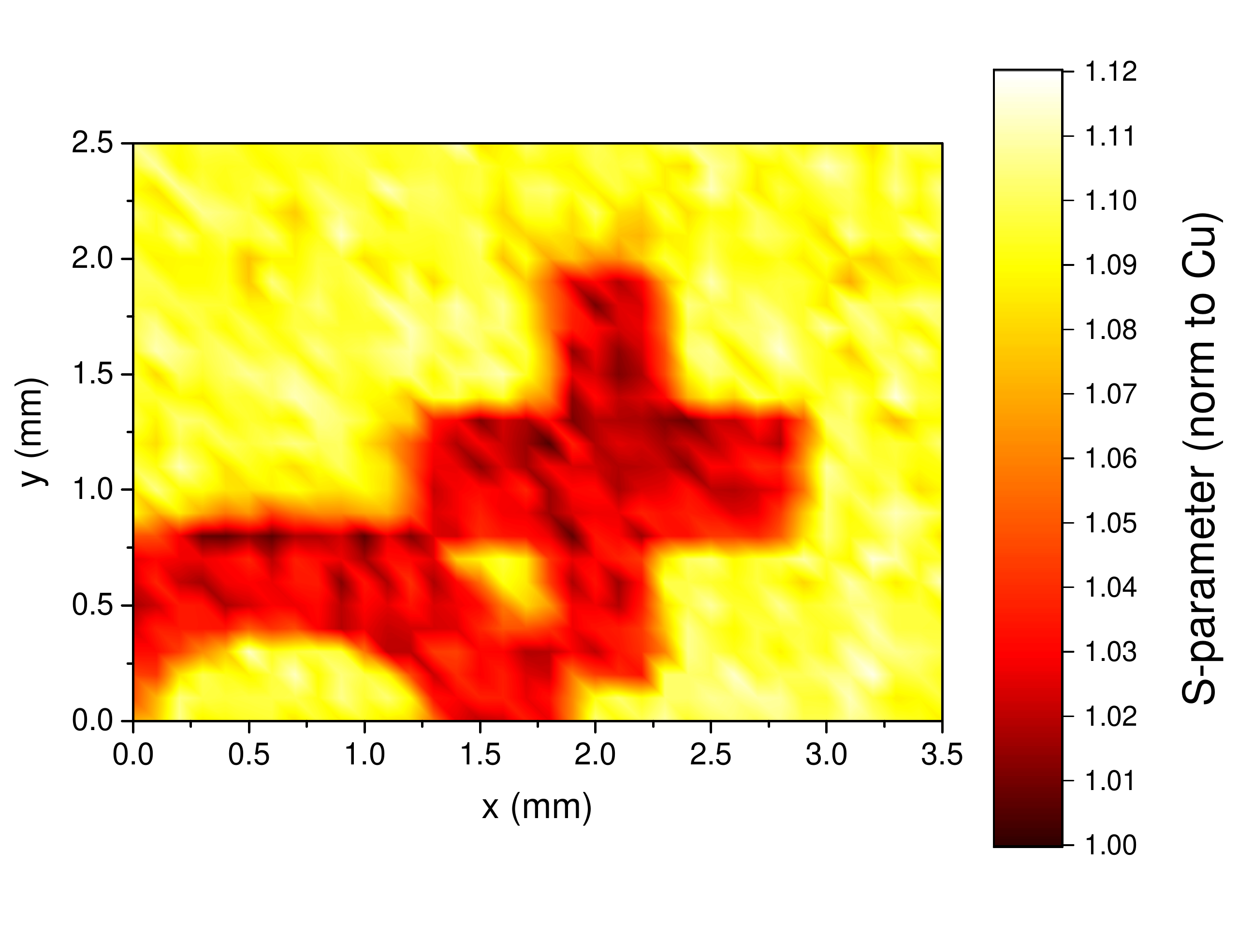}}
    \end{minipage}
	\caption{2D S-parameter maps of Cu \textbf{e$^{+}$ }patterns etched in a circuit board. 
	(a) Optical image of the sample. The scan areas of the positron beam are marked (red rectangles).
	Scan using the NEPOMUC remoderated beam with $\Delta_{x,y}$\,=\,200\,$\mu$m of the (b) largest (t$_m$\,=\,13\,s) and (c) the smallest \textbf{e$^{+}$} pattern (t$_m$\,=\,7\,s), 
	(d) detail of the smallest pattern mapped with the positron microbeam  providing the best spatial resolution ($\Delta_{x,y}$\,=\,100\,$\mu$m, t$_m$\,=\,25\,s).}
	\label{eplusses}
\end{figure}

The performance for both beam settings, i.e. for the NEPOMUC remoderated beam and the positron microbeam, was tested on two specially patterned samples by 2D scanning with various step widths ($\Delta_{x}, \Delta_{y}$) and measurement time (t$_m$).
First, a sample consisting of a Cu mesh with 50\,$\mu$m thick bars separated by 204\,$\mu$m, mounted on an Al sample holder was scanned in x- and y-direction ($\Delta_{x,y}$\,=\,5$\mu$m, t$_m$\,=7\,s) yielding a 2D S-parameter map as shown in figure\,\ref{Mesh100}b.
It could be demonstrated that, despite the low measurement time per point  of only 7\,s, the Cu bars can be resolved, and, more importantly, the 2D map shows no distortions.
Then 2D scans of the S-parameter have been performed on a sample with "e$^{+}$"-patterns of different size etched in an electronic circuit board (figure\,\ref{eplusses}a). 
The S-parameter was obtained by averaging the data of all four Ge detectors.
The energy was set to 30\,keV and 25\,keV for the NEPOMUC remoderated beam and the positron microbeam, respectively.
Due to the high contrast in the first image (figure\,\ref{eplusses}b) recorded with the NEPOMUC remoderated beam ($\Delta_{x,y}$\,=\,200$\mu$m, t$_m$\,=\,13\,s) t$_m$ was reduced to 7\,s for the scan of the smallest e$^{+}$ (figure\,\ref{eplusses}c).
Despite blurring the contours, the NEPOMUC remoderated beam with 250\,$\mu$m diameter is able to resolve the image of the smallest e$^{+}$ as well. 
Note that no distortions are observed over the complete range of the scan area. 
However, using the positron microbeam with a resolution of 50\,$\mu$m (FWHM), $\Delta_{x,y}$\,=\,100$\mu$m and  t$_m$\,=\,25\,s the pattern can be resolved without any blurring and distortions also for large areas of up to 19$\times$19\,mm$^2$.
Consequently, the new spectrometer provides outstanding performance in routine operation enabling high resolution 2D defect imaging  with unprecedented short measurement times of about 160\,min/mm$^2$ with the positron microbeam ($\Delta_{x,y}$\,=\,50\,$\mu$m, t$_m$\,=\,25\,s per point) and $<$\,2\,min/mm$^2$ with the NEPOMUC remoderated beam ($\Delta_{x,y}$\,=\,250\,$\mu$m, t$_m$\,=\,7\,s).

In order to yield the highest remoderation efficiency, $\epsilon_{rem}$, the Ni(100) foil was first heated to 500\,$^{o}$C in UHV for one hour. 
Then it was held at the same temperature for an additional hour in a H atmosphere of 10$^{-3}$\,mbar.  
The remoderation efficiency $\epsilon_{rem}$ is calculated from intensity measurements at different conditions with respect to the intensity of the NEPOMUC remoderated beam with an intensity of typically $3.0\cdot 10^7 $ moderated positrons per second (see table\,\ref{tab:eff_ni}).  
This value compares well with the measured count rate of 60000\,cps in a single HPGe detector of the CDB spectrometer taking into account the solid angle and the detector efficiencies. 
Inserting the Ni(100) foil in the as-received state reduces the count rate to 510\,cps yielding only $\epsilon_{rem}$\,=\,0.85\,\%. 
The increase to $\epsilon_{rem}$\,=\,3.5\,\% after heating is attributed to the desorption of the surface impurities  to some extent in accordance with the temperature dependent XPS measurements presented in figure\,\ref{XPS_Ni}.
After heat treatment in a H atmosphere, a count rate of 5800\,cps could be achieved resulting in a remoderation efficiency of 9.6\,\%. 

\begin{table}[h!]
\centering
\caption{Remoderation efficiency $\epsilon_{rem}$ of the brightness enhancement system of the positron microbeam.}
\label{tab:eff_ni}
\vspace{0,25cm}
\begin{tabular}{|l|c|c|}
\hline 
\textbf{Condition of Ni(100) foil} & \textbf{Count rate} & \textbf{$\epsilon_{rem}$} \\
 & (cps) & (\%)\\ 
\hline \hline
removed & 60000 & -- \\ 
\hline 
as received & 510 & 0.85 \\ 
\hline 
heated to 500$^{o}$C in UHV & 2100 & 3.5 \\ 
\hline 
heated to 500$^{o}$C + H$_2$ & 5800 & 9.6 \\ 
\hline 
\end{tabular}
\end{table}

%++++++++++++++++++++++++++++++++++++++++++++++++++++++++++++++++++++++++++++++++++
\section{Defects and precipitates in laser beam welded AlCu6Mn}
%++++++++++++++++++++++++++++++++++++++++++++++++++++++++++++++++++++++++++++++++++
\subsection{Laser beam welding}
%++++++++++++++++++++++++++++++++++++++++++++++++++++++++++++++++++++++++++++++++++
Laser beam welding (LBW) is an excellent technique that frequently outperforms traditional arc welding processes due to various advantages. 
The laser beam can typically be focused in the micrometer range with high accuracy and reproducibility.
Hence, complicated joint geometries can be welded with high precision. 
The high welding speed in the range of 100\,mm/s and the small laser spot, which introduces just a small amount of heat, result in minor changes of the micro structure and low thermal distortions. 
This in turn leads to a more narrow heat affected zone (HAZ) compared to other welding techniques.
In addition, cavity-free welds can be realized with high reliability enhancing the mechanical strength of the joint \cite{WeldTech2013}.
Especially welding of high-strength Al alloys, containing Cu or Li, is a key technology in modern manufacturing engineering \cite{Dit_Al_LBW2011} since lightweight constructions more and more replace heavier steel and riveted joint constructions \cite{Dit_Al_LBW2011,Ost07_Al}.
Therefore, in order to avoid deterioration of  the mechanical properties it is of particular interest to produce high-strength welds with low defect concentration. 

In the present work, we focus on the characterization of a laser beam welded Al alloy (AlCu6Mn, EN AW-2219 T87) with a Cu content of 6.3\,\%. 
This alloy is a typical example of a precipitation hardened material.
In order to significantly enhance the strength of the material, Cu atoms are dissolved in the crystal lattice during a first heat treatment, and subsequently, artificial aging leads to the formation of Cu precipitates \cite{WeldKou2003}.
Due to the inhomogeneous temperature distribution, LBW of this alloy is expected to lead to various position dependent effects such as formation and quenching of defects as well as the dissolution of Cu precipitates.  
This in turn results in a local variation of the mechanical strength in the weld and in the HAZ.

%++++++++++++++++++++++++++++++++++++++++++++++++++++++++++++++++++++++++++++++++++
\subsection{Defect imaging of a laser beam weld}
\begin{figure}

	\centering
  \includegraphics[width=\textwidth]{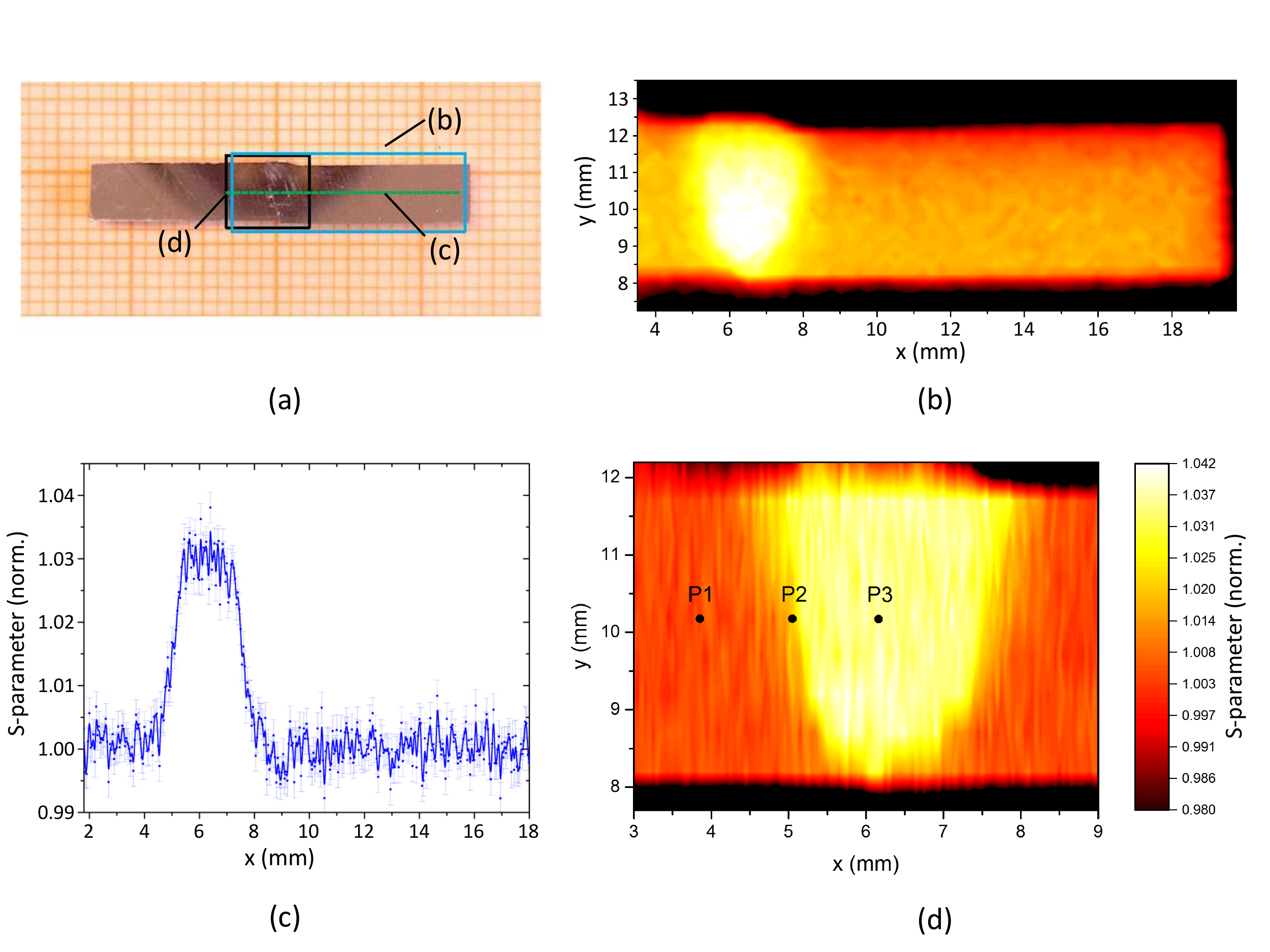}
		\caption{Laser beam weld (LBW) of the Al alloy EN AW-2219 T87: (a) Optical image of the cross cut of the sample. The scan areas are highlighted (blue and black rectangles, green line). 
(b) 2D map recorded with the NEPOMUC remoderated  beam ($\Delta_{x,y}$\,=\,200\,$\mu$m),
(c) High resolution line scan of the LBW at y\,=\,10\,mm ($\Delta_x$\,=\,50\,$\mu$m) with the CDB positron microbeam. The steep changes of the S-parameter within $<$1\,mm  indicate a very small heat affected zone.
(d) High resolution 2D map obtained with the CDB microbeam ($\Delta_{x}$\,=\,50\,$\mu$m, $\Delta_{y}$\,=\,500\,$\mu$m). The points P1-P3 mark the positions of additional CDBS measurements.}
	\label{LBW_image}
\end{figure}

%++++++++++++++++++++++++++++++++++++++++++++++++++++++++++++++++++++++++++++++++++
We apply (C)DB spectroscopy with the positron microbeam in order to image the concentration of open volume defects and to detect the local dependent formation of Cu precipitates.
A laser beam weld of two Al sheets (EN AW-2219 T87, $\rho\,=\,2.85\,g/cm^3$) with a thickness of 4\,mm using a single-mode laser (IPG YLR-3000, spot size:\;50\,$\mu$m, welding speed:\;100\,mm/s) was produced at the Institute for Machine Tools and Industrial Management (IWB) at TUM. 
The polished cross cut of the sample is shown in figure\,\ref{LBW_image}\,a.

First, an overview 2D map of the S-parameter (figure\,\ref{LBW_image}b) was recorded using the NEPOMUC remoderated  beam ($\Delta_{x,y}$\,=\,200\,$\mu$m) with an energy of 30\,keV according to a mean positron implantation depth of 3.3\,$\mu$m.
Compared to the not affected material the S-parameter is drastically increased in the region of the laser beam weld that is attributed to the creation and quenching of a large amount of vacancy-like defects during the welding process. 
Note that a gradient of the S-parameter is observed in y direction of the Al sample outside of the welded zone. 
This effect in the as-received material can be explained by more vacancy-like defects on one side (higher S-parameter at the bottom side) generated during cold-rolling of the metal sheets. 

The high resolution line scan recorded with the positron microbeam across the weld  at a fixed y\,=\,10\,mm and with $\Delta_{x}$\,=\,50\,$\mu$m is shown figure\,\ref{LBW_image}\,c.
The abrupt changes of the S-parameter at around x\,=\,5 and 7.5\,mm indicate a small HAZ of $<$\,1\,mm.
The area around the welding zone was also examined by a 2D scan ($\Delta_{x}$\,=\,500\,$\mu$m  and $\Delta_{y}$\,=\,50\,$\mu$m) as shown in figure\,\ref{LBW_image}\,d. 
A very sharp transition between the weld and the surrounding material also confirms the effect of a well localized HAZ.

%++++++++++++++++++++++++++++++++++++++++++++++++++++++++++++++++++++++++++++++++++
\subsection{Detection of Cu precipitates in laser beam welded AlCu6Mn}
%++++++++++++++++++++++++++++++++++++++++++++++++++++++++++++++++++++++++++++++++++
Bulk CDB measurements were performed  in the middle, at the edge and outside the weld (see points P1-P3 in figure\,\ref{LBW_image}\,d) in order to get a deeper insight in the type of defects and to obtain information on the elements probed by the positrons. 
The analysis of the CDB spectra was performed as described in \cite{Rei16a}.

\begin{figure}
	\centering
	\includegraphics[width=0.7\textwidth]{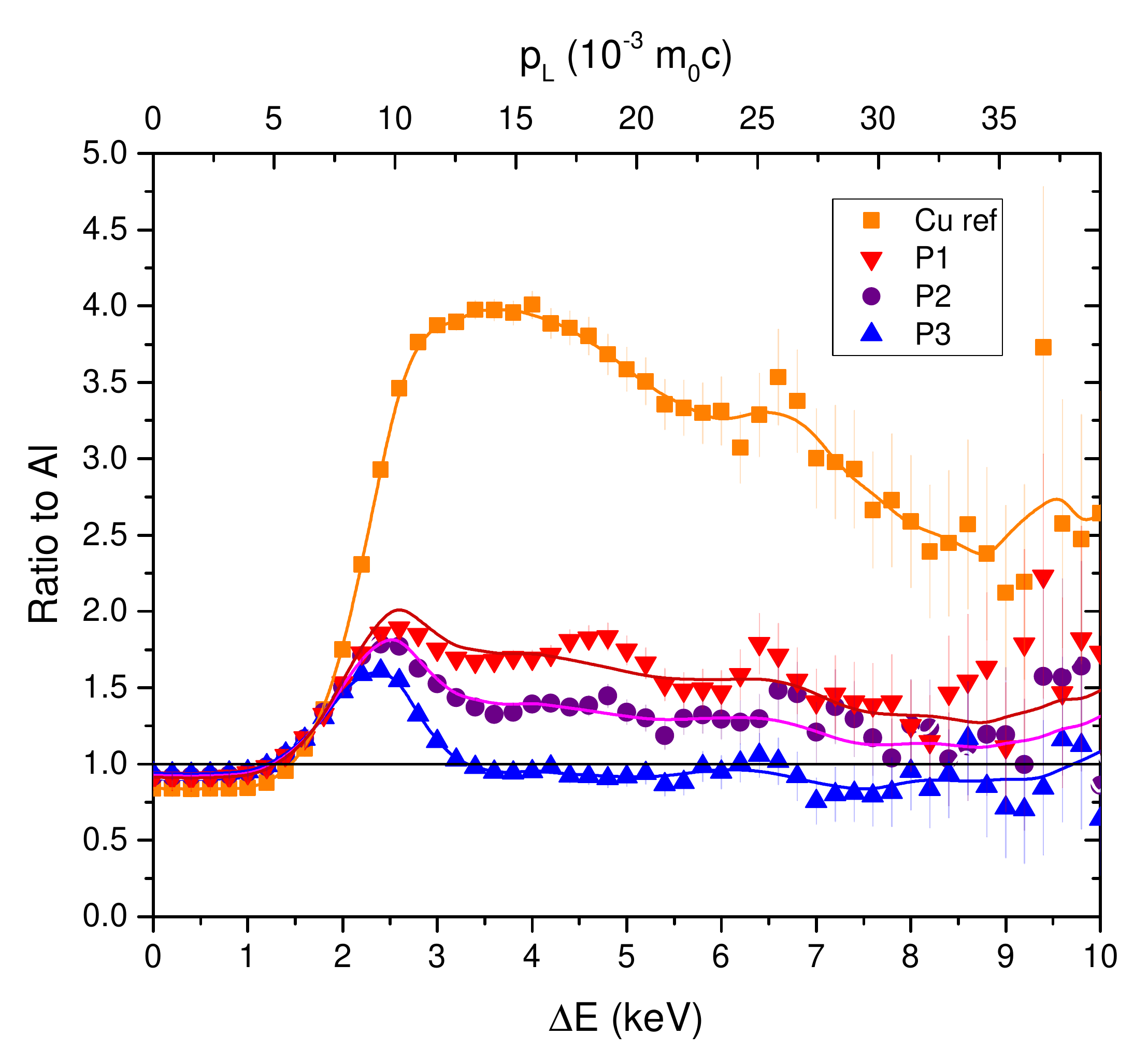}
    \caption{CDB ratio curves with respect to pure Al recorded with the positron microbeam at positions P1-P3 as marked in figure\,\ref{LBW_image} and for pure Cu as reference. 
The so-called confinement peak at 9$\cdot\,10^{-3}\,m_0c$ at P3 clearly indicates the presence of vacancies inside the weld.
The observed Cu signature at P1 and P2 is attributed to Cu rich precipitates in the as-received material and in the heat affected zone.
The fraction of annihilation events at Cu is obtained by a weighted fit of the Cu reference spectrum and the spectrum obtained at P3 (solid lines) yielding Cu intensities of 25.7\,\% and 14.8\,\%  for P1 and P2, respectively.
 	}
	\label{CDB_LBW}
\end{figure}

Figure\,\ref{CDB_LBW} shows so-called ratio curves (with respect to pure Al) of the annihilation line at three selected points (shown in figure\,\ref{LBW_image}\,c) and for pure Cu as reference.
The signature of Cu can be seen in the spectra obtained at P1 and P2 with different intensities but not at all in the center of the weld (P3). 
The pronounced peak at around 9$\cdot\,10^{-3}\,m_0c$ in the ratio curve at P3 clearly indicates the presence of vacancies in the alloy.
Such vacancies with a positron trapping potential in Al of 1.75\,eV \cite{Xia90} lead to quantum confinement  of the positron wave function resulting in a non-negligible momentum.
The measured position of this so-called confinement peak is in excellent agreement with CDB spectra of vacancies in Al produced after deformation and quenching \cite{Cal05}. Similar effects are also observed for supersaturated solid solution of as-quenched \cite{Som02} and aged AlCu alloys \cite{Fol07}.
The fact that no Cu signal is observed is explained by the less favourable, if at all, formation of vacancy-Cu complexes inside the weld.

The relative fraction of annihilation events with electrons from Cu at P1 and P2 are estimated by a superposition of the Cu reference spectrum and the spectrum obtained at P3 (for the applied analysis method see e.g. \cite{Nag00,Hug08a,Rei14a}). 
The latter spectrum is chosen in order to account for the presence of vacancies. A best fit to the data (see figure\,\ref{CDB_LBW}) yields Cu intensities of 25.7\,\% and 14.8\,\%  for P1 and P2, respectively.

The significant contribution of the Cu signature in the as-received material (P1) and to lesser extent in the HAZ of the weld (P2) is attributed to positron trapping at Cu rich precipitates. 
This observation is in agreement with artificially age-hardened precipitates formed in the $\Theta$ phase, i.e. Al$_2$Cu phase, which is incoherent with the lattice of the solid solution.
Hence, both the higher positron affinity of Cu compared to Al ($A^+_{Cu}\,=\,-4.81$ and $A^+_{Al}\,=\,-4.41$ \cite{Pus89}) as well as lattice distortions at the interface of the precipitates and the Al matrix  \cite{WeldKou2003} lead to a higher positron annihilation rate with electrons from Cu.

Note that a single Cu atom solved in an Al matrix cannot lead to positron trapping.\footnote{Applying a quantum-well model the minimum radius of a Cu cluster embedded in Al for confining the positron wave function can be estimated by 
$r_c \cong 5.8\cdot a_0/\sqrt{\left|A^+_{Cu}-A^+_{Al}\right|/(eV)}$ \cite{Pus89} (with Bohr radius a$_0$) yielding  $r_c\,=\,0.49\,nm$. } 
Consequently, inside the weld the disappearance of the Cu signature (P3 in figure\,\ref{CDB_LBW} d) is explained by melting, i.e. dissolution of the Cu rich phases, and rapid cooling in the welded zone where Cu atoms are kinetically hindered to form precipitates and hence form a supersaturated solid solution \cite{WeldKou2003}.
This effect is expected to be enhanced for laser beam welded joints since the small spot size and the high welding speed of the laser as well as the high thermal conductivity of the material of 120\,W/mK \cite{Al_ASM93} cause a high heat flow away from the weld.

%++++++++++++++++++++++++++++++++++++++++++++++++++++++++++++++++++++++++++++++++++
\section{Conclusion}
%++++++++++++++++++++++++++++++++++++++++++++++++++++++++++++++++++++++++++++++++++

A new positron microbeam for the investigation and 2D imaging of defects on an atomic scale using (coincindent) DB spectroscopy has been successfully put into operation. 
For this purpose we realized a \textit{threefold moderated} monoenergetic positron beam and upgraded the CDB spectrometer located at the high intensity positron source NEPOMUC.
By introducing a transmission remoderator, consisting of a 100\,nm thick Ni(100) foil, for brightness enhancement, a diameter as low as 33\,$\mu$m could be achieved.
Benefiting from the high beam intensity of NEPOMUC and the achieved remoderation efficiency of the new transmission remoderator (9.6\,\%), the area of the beam spot could be reduced by a factor of 60. It was demonstrated that large samples can be examined with high lateral resolution and without distortions over the complete scanning area of 19$\times$19\,mm.
Apart from the new remoderation unit several improvements have been made to facilitate the operation and to enhance the robustness of the positron beam.  
For quick measurements without additional remoderator a beam diameter of about 250\,$\mu$m is achieved which can be used in combination with different insertions for the sample environment enabling temperature dependent measurements in the range of 40\,K--1000\,K.
Most importantly, the new CDB spectrometer provides excellent performance in routine operation for high resolution 2D defect imaging  with unprecedented short measurement times  of about 160\,min/mm$^2$ with the positron microbeam ($\Delta_{x,y}$\,=\,50\,$\mu$m, t$_m$\,=\,25\,s per point) and $<$2\,min/mm$^2$ with the NEPOMUC remoderated beam ($\Delta_{x,y}\,=$\,250\,$\mu$m, t$_m$\,=\,7\,s). 3D imaging of defects can be performed by varying the positron beam energy.

For the first time, high resolution experiments were performed to study both, the defect distribution and the presence of Cu precipitates, showing the outstanding capability of the new CDB spectrometer.
Since laser beam welding yields joints of small lateral extent a positron microbeam is highly beneficial for the application of spatial resolved defect spectroscopy.
In our experiment, the examination of a laser beam weld of the high-strength Al alloy (AlCu6Mn, EN AW-2219 T87) revealed a sharp transition between the raw material and the welded zone as well as a very small heat affected zone. 
Besides vacancy-like defects, Cu rich precipitates could be identified in the as-received material and to a lesser extent in the transition zone of the weld. 
Most notably, inside the weld we could clearly identify the dissolution of the Cu atoms in the crystal lattice, i.e.\,formation of a supersaturated solution, as well as the presence of vacancies without forming Cu-vacancy complexes.

In the future, we aim to to study selected sets of welds produced by different welding techniques and  various welding parameters using the positron microbeam now available at the upgraded CDB spectrometer as a powerful tool for  high resolution defect spectroscopy.
In particular, it is expected to gain an improved understanding of the mechanical properties of welds by combining complementary techniques such as hardness measurements and  optical microscopy with spatially resolved (C)DB spectroscopy.

%++++++++++++++++++++++++++++++++++++++++++++++++++++++++++++++++++++++++++++++++++
\section*{Acknowledgements}
%++++++++++++++++++++++++++++++++++++++++++++++++++++++++++++++++++++++++++++++++++

We would like to thank M.Sc.\,A. Bachmann from IWB/TUM for providing LBW samples of the technical Al alloys.
The authors also thank Prof.\,M. Fujinami from Chiba university and Dr.\,N. Oshima from AIST for fruitful discussions.
Financial support by the German federal ministry of education and research (BMBF) within the Project No. 05K13WO1 is gratefully acknowledged.

\textit{Authors contribution:}
T.G. and C.H. wrote the manuscript. T.G. designed the new spectrometer. C.H. initiated and managed the project. T.G., L.B., M.T. and S.V. set up the new instrument. 
M.D. and  B.R. improved the software and the beam control.

\newcommand{\newblock}{}
\bibliographystyle{unsrt}

\end{document}